\pdfoutput=1
\pdfoutput=1
\pdfoutput=1
\pdfoutput=1
\pdfoutput=1
\RequirePackage{fix-cm}
\documentclass[twocolumn,epjc3-upd,fleqn]{svjour3}
\smartqed  % flush right qed marks, e.g. at end of proof
\RequirePackage{graphicx}
\RequirePackage{subfigure}
\usepackage{rotating}
\usepackage{amsmath}
\usepackage{mathtools}
\usepackage{multirow}
\RequirePackage[colorlinks,citecolor=blue,urlcolor=blue,linkcolor=blue]{hyperref}
\usepackage{doi}
\usepackage[]{units}
\usepackage{xspace}
\usepackage[switch]{lineno} 
\journalname{Eur. Phys. J. C}
\def\exposure{9.95~kg~yr\xspace}
\def\kgyr{kg~yr\xspace}

\def\cuz{\mbox{CUPID-0}\xspace}
\def\qz{\mbox{CUORE-0}\xspace}

\def\pbext{\emph{ExtPb}\xspace} 
\def\cryoext{\emph{CryoExt}\xspace} 
\def\cryoint{\emph{CryoInt}\xspace} 
\def\pbint{\emph{IntPb}\xspace} 
\def\holder{\emph{Holder}\xspace}

\def\crystals{\emph{Crystals}\xspace}
\def\reflectors{\emph{Reflectors}\xspace}

\def\monealfa{$\mathcal{M}_{1\alpha}$\xspace}
\def\monebeta{$\mathcal{M}_{1\beta/\gamma}$\xspace}
\def\mtwo{$\mathcal{M}_{2}$\xspace}
\def\mtwosum{$\Sigma_{2}$\xspace}

\def\u{$^{238}$U\xspace}
\def\th{$^{232}$Th\xspace}

\def\acddo{$^{228}$Ac\xspace}

\def\biduq{$^{214}$Bi\xspace}

\def\podd{$^{210}$Po\xspace}

\def\tld{$^{208}$Tl\xspace}

\def\seod{$^{82}$Se\xspace}

\def\cosz{$^{60}$Co\xspace}

\def\mncq{$^{54}$Mn\xspace}

\def\kq{$^{40}$K\xspace}

\def\znsc{$^{65}$Zn\xspace}
\def\udtc{$^{235}$U\xspace}

\def\znsenr{Zn$^{82}$Se\xspace}

\def\bb{$\beta\beta$\xspace}
\def\bbd{$2\nu\beta\beta$\xspace}
\def\bbz{$0\nu\beta\beta$\xspace}

\def\rate{counts/(keV$\cdot$kg$\cdot$y)\xspace}
\def\vita-dim{$T_{1/2}$\xspace}

\def\be{\begin{equation}}
\def\ee{\end{equation}}

\def\rate{counts/(keV~kg~yr)\xspace}

\begin{document}
%\linenumbers
\title{Background Model of the CUPID-0 Experiment}
\author{O.~Azzolini\thanksref{Legnaro}
\and J.~W.~Beeman\thanksref{LBNL}
\and F.~Bellini\thanksref{Roma,INFNRoma}
\and M.~Beretta\thanksref{MIB,INFNMiB}
\and M.~Biassoni\thanksref{INFNMiB}
\and C.~Brofferio\thanksref{MIB,INFNMiB}
\and C.~Bucci\thanksref{LNGS}
\and S.~Capelli\thanksref{MIB,INFNMiB}
\and L.~Cardani\thanksref{INFNRoma}
\and P.~Carniti\thanksref{MIB,INFNMiB}
\and N.~Casali\thanksref{INFNRoma}
\and D.~Chiesa\thanksref{MIB,INFNMiB,e1}
\and M.~Clemenza\thanksref{MIB,INFNMiB}
\and O.~Cremonesi\thanksref{INFNMiB}
\and A.~Cruciani\thanksref{Roma,INFNRoma}
\and I.~Dafinei\thanksref{INFNRoma}
\and S.~Di~Domizio\thanksref{Genova,INFNGenova}
\and F.~Ferroni\thanksref{INFNRoma,GSSI}
\and L.~Gironi\thanksref{MIB,INFNMiB}
\and A.~Giuliani\thanksref{CNRS}
\and P.~Gorla\thanksref{LNGS}
\and C.~Gotti\thanksref{MIB,INFNMiB}
\and G.~Keppel\thanksref{Legnaro}
\and M.~Martinez\thanksref{Roma,INFNRoma,e4}
\and S.~Nagorny\thanksref{LNGS,GSSI,e2}
\and M.~Nastasi\thanksref{MIB,INFNMiB}
\and S.~Nisi\thanksref{LNGS}
\and C.~Nones\thanksref{CEA}
\and D.~Orlandi\thanksref{LNGS}
\and L.~Pagnanini\thanksref{MIB,INFNMiB}
\and M.~Pallavicini\thanksref{Genova,INFNGenova}
\and L.~Pattavina\thanksref{LNGS}
\and M.~Pavan\thanksref{MIB,INFNMiB}
\and G.~Pessina\thanksref{INFNMiB}
\and V.~Pettinacci\thanksref{INFNRoma}
\and S.~Pirro\thanksref{LNGS}
\and S.~Pozzi\thanksref{MIB,INFNMiB}
\and E.~Previtali\thanksref{MIB,INFNMiB}
\and A.~Puiu\thanksref{MIB,INFNMiB}
\and C.~Rusconi\thanksref{LNGS,USC}
\and K.~Sch\"affner\thanksref{LNGS,GSSI}
\and C.~Tomei\thanksref{INFNRoma}
\and M.~Vignati\thanksref{INFNRoma}
\and A.~Zolotarova\thanksref{CEA, e3}
}
\institute{INFN - Laboratori Nazionali di Legnaro, Legnaro (Padova) I-35020 - Italy \label{Legnaro}
\and
Materials Science Division, Lawrence Berkeley National Laboratory, Berkeley, CA 94720 - USA\label{LBNL}
\and
Dipartimento di Fisica, Sapienza Universit\`{a} di Roma, Roma I-00185 - Italy \label{Roma}
\and
INFN - Sezione di Roma, Roma I-00185 - Italy\label{INFNRoma}
\and
Dipartimento di Fisica, Universit\`{a} di Milano - Bicocca, Milano I-20126 - Italy\label{MIB}
\and
INFN - Sezione di Milano - Bicocca, Milano I-20126 - Italy\label{INFNMiB}
\and
INFN - Laboratori Nazionali del Gran Sasso, Assergi (L'Aquila) I-67100 - Italy\label{LNGS}
\and
Gran Sasso Science Institute, 67100, L'Aquila - Italy\label{GSSI}
\and
Dipartimento di Fisica, Universit\`{a} di Genova, Genova I-16146 - Italy\label{Genova}
\and
INFN - Sezione di Genova, Genova I-16146 - Italy\label{INFNGenova}
\and
CSNSM, Univ. Paris-Sud, CNRS/IN2P3, Universit\'e Paris-Saclay, 91405 Orsay - France\label{CNRS}
\and
IRFU, CEA, Universit\'e Paris-Saclay, F-91191 Gif-sur-Yvette, France\label{CEA}
\and
Department of Physics  and Astronomy, University of South Carolina, Columbia, SC 29208 - USA\label{USC}
}
\thankstext{e1}{Corresponding Author: davide.chiesa@mib.infn.it}
\thankstext{e4}{Present Address: Fundaci\'on ARAID and Laboratorio de F\'isica Nuclear y Astropart\'iculas, Universidad de Zaragoza, C/ Pedro Cerbuna 12, 50009 Zaragoza, Spain}
\thankstext{e2}{Present Address: Queen's University, Physics Department, K7L 3N6, Kingston (ON), Canada}
\thankstext{e3}{Present Address: CSNSM, Univ. Paris-Sud, CNRS/IN2P3, Universit\'e Paris-Saclay, 91405 Orsay, France}

\date{Received: 24 April 2019 / Accepted: 27 June 2019}
% The correct dates will be entered by the editor
\twocolumn
\maketitle
\begin{abstract}
\cuz is the first large mass array of enriched \znsenr scintillating low temperature calorimeters, operated at LNGS since 2017.
During its first scientific runs, \cuz collected an exposure of 9.95~\kgyr.
Thanks to the excellent rejection of $\alpha$ particles, we attained the lowest background ever measured with thermal detectors in the energy region where we search for the signature of $^{82}$Se neutrinoless double beta decay.
In this work we develop a model to reconstruct the \cuz background over the whole energy range of experimental data. We identify the background sources exploiting their distinctive signatures and we assess their extremely low contribution (down to $\sim10^{-4}$ counts/ (keV~kg~yr)) in the region of interest for $^{82}$Se neutrinoless double beta decay search. This result represents a crucial step towards the comprehension of the background in experiments based on scintillating calorimeters and in next generation projects such as CUPID.

\keywords{double beta decay \and scintillating calorimeters \and background model}
\PACS{23.40.-s $\beta$ decay; double $\beta$ decay; electron and muon capture \and 27.50.+e mass 59 $\leq$ A $\leq$ 89 \and 
29.30.Kv X- and $\gamma$-ray spectroscopy}
\end{abstract}

\section{Introduction}
\label{sec:introduction}
The postulated neutrinoless double beta decay (\bbz) consists of two neutrons of an atomic nucleus simultaneously decaying to two protons and two electrons, without the accompanying emission of electron antineutrinos \cite{Furry:1939qr}.
If observed, \bbz would provide crucial evidence for lepton number violation and it is one of the most sensitive methods to study neutrino properties such as its nature (Dirac or Majorana) and the absolute value of its mass \cite{DellOro:2016tmg}.
The experimental signature of \bbz is a peak at the end of the continuous spectrum produced by the electrons emitted in two-neutrino double beta decay, an allowed, although extremely rare, second order nuclear transition.
Detectors with excellent energy resolution, such as low temperature calorimeters (historically also called \textit{bolometers}), are the best candidates to study this process, being able to disentangle the searched peak from the continuous background.
However, the energy resolution is only one of the parameters that concur to determine the sensitivity of an experiment for the search of \bbz decay. Others are the number of isotopes under study, the live time, the detection efficiency and the background rate in the energy region of interest (ROI)~\cite{Alduino:2017pni_Sensitivity}.

The goal of the background model described in this work is to identify the sources of the \cuz background and evaluate their contribution to the ROI around the \seod \bbz Q-value  (2997.9$\pm$0.3~keV~\cite{Lincoln:2012fq}).
This study is, in particular, fundamental for the design of next generation experiments, because the conventional techniques applied to measure radioactivity in materials are not able to probe levels of contamination as low as those required for future \bbz experiments. 
Therefore the required information must be extrapolated from current rare event experiments.

In this paper, after introducing the \cuz detector and the data production (Section~\ref{sec:exp}), we analyze the experimental spectra in wide energy ranges, from a few hundred keV to $\sim$10~MeV, to find signatures of background sources (Section~\ref{sec:BkgAnalysis}). The energy spectra produced in the detector by each source are then simulated by means of a Monte Carlo code (Section~\ref{sec:MC}). 
The background model (Section~\ref{sec:BkgModel}) is constructed by selecting a representative list of sources whose spectra are combined in a Bayesian fit to the experimental data. Information about contaminant activities available from independent measurements or analyses are included through apposite \textit{prior} distributions.
Finally (Section~\ref{sec:Results}), we present the fit results, i.e. the activities obtained for the background sources and their contribution to the \bbz ROI, as well as a discussion of systematic uncertainties.

\section{CUPID-0: detector and data production}
\label{sec:exp}
\subsection{The \cuz detector}
The detection technique used in \cuz experiment~\cite{Azzolini:2018dyb_C0PRL} is based on cryogenic scintillating calorimeters. These devices allow a simultaneous detection of energy released as heat and light. We exploit both signals to identify different types of interacting particles.
The \cuz detector is a five tower array of 26 ZnSe scintillating crystals, 24 enriched in $^{82}$Se at 95\% level and 2 with $^{82}$Se natural isotopic abundance. The total detector mass is 10.5~kg of ZnSe, equivalent to 5.17~kg of $^{82}$Se.
The crystals are interleaved with Germanium Light Detectors (Ge-LDs), that are used to measure the scintillation signal produced in ZnSe by interacting particles.
Both ZnSe crystals and Ge-LDs are held in position by means of PTFE clamps and are thermally coupled to a heat bath at $\sim$10~mK by means of a copper structure.
In order to increase the light collection, each ZnSe crystal is surrounded by a VIKUITI\textsuperscript{TM} multi-layers reflecting foil produced by 3M.
\cuz is hosted in Hall A of Laboratori Nazionali del Gran Sasso (LNGS), inside the cryostat previously used for the CUORICINO and \qz experiments~\cite{Andreotti:2010vj,Alfonso:2015wka_Q0PRL}. The shielding infrastructure is identical to the \qz one, with the only difference that the 10~mK thermal shield has not been installed, and that the masses of 50~mK and 600~mK shields have been reduced via electrical discharge machining (EDM). 
The radio-purity of materials used in \cuz experimental setup has been measured with different techniques~\cite{Beeman:2015xjv,Azzolini:2018tum_C0det,Alduino:2016vtd_Q02nbb,Alduino:2016vjd_Q0det,BiPoDetector}, obtaining the results reported in Tab.~\ref{tab:Limits}.

\begin{table}[b!]
\begin{center}
\caption{Measurements and limits on contaminations of \cuz detector components. Ge-LD radiopurity is certified by UMICORE company. The contaminations of VIKUITI\textsuperscript{TM} foils have been measured via Inductively Coupled Plasma Mass Spectrometry (ICPMS) at LNGS~\cite{Azzolini:2018tum_C0det} and with a BiPo detector~\cite{BiPoDetector} (private communication). The limits of the other components are taken from Ref.~\cite{Alduino:2016vtd_Q02nbb}. Error bars are 1~$\sigma$, limits are 90\% C.L. upper limits.}
\begin{tabular}{ccc}
\hline
Component                & $^{232}$Th                  & $^{238}$U\\
                         & [Bq/kg]                     & [Bq/kg] \\
\noalign{\smallskip}\hline
Ge-LD                 & $<$6$\times$10$^{-6}$     &  $<$1.9$\times$10$^{-5}$       						\\
\multirow{2}{*}{VIKUITI\textsuperscript{TM}}               & (4.9 $\pm$ 1.2$)\times$10$^{-5}$ &
(1.7$\pm$0.5)$\times$10$^{-4}$ \\
& $<$8.4 $\times$10$^{-5}$~($^a$) & -\\
Epoxy Glue               & $<$8.9$\times$10$^{-4}$     &  $<$1.0$\times$10$^{-2}$ \\
Au bonding wires         & $<$4.1$\times$10$^{-2}$     & $<$1.2$\times$10$^{-2}$       				\\
Si heaters            & $<$3.3$\times$10$^{-4}$     &  $<$2.1$\times$10$^{-3}$       						\\
Ge thermistors          & $<$4.1$\times$10$^{-3}$     &  $<$1.2$\times$10$^{-2}$         					\\
PTFE supports           & $<$6.1$\times$10$^{-6}$     &  $<$2.2$\times$10$^{-5}$       						\\
Cu NOSV                 & $<$2.0$\times$10$^{-6}$     &  $<$6.5$\times$10$^{-5}$     \\
\noalign{\smallskip}\hline
\multicolumn{3}{l}{($^a$) Limit on $^{212}$Bi--$^{212}$Po contamination.} \\
\end{tabular}
\label{tab:Limits}
\end{center}
\end{table}

Both ZnSe crystals and Ge-LDs are equipped with a Neutron Transmutation Doped (NTD) Ge thermistor \cite{Wang:1989vk}, working as temperature-voltage transducer.
A P-doped Si Joule heater~\cite{Andreotti:2012zz,Carniti:2017zkr}, glued to each device, periodically injects a constant energy reference pulse used to measure gain variations induced by temperature fluctuations.
The front-end electronics comprises an amplification stage, a six-pole anti-aliasing active Bessel filter and an 18 bits ADC board~\cite{Arnaboldi:2015wvc,Arnaboldi:2017aek}.
The complete data-stream is digitized with a frequency of 1~kHz (2~kHz) for ZnSe (Ge-LD) and saved on disk in NTuples based on the ROOT software framework~\cite{DiDomizio:2018ldc}. 
A software derivative trigger with channel dependent threshold is implemented online. 
When a trigger fires on a crystal, the waveforms of the corresponding Ge-LDs are also flagged as signals.
For each event on ZnSe we analyze a window of \unit[5]{s} (\unit[1]{s} before the trigger and \unit[4]{s} after it). 
The analysis window of signals on Ge-LDs is \unit[500]{ms} long (\unit[100]{ms} before the trigger and \unit[400]{ms} after it).
The samples before the trigger provide the baseline temperature of the detector, while the remaining samples are used to determine the pulse amplitude and shape, for evaluating the deposited energy. 
More details about the \cuz detector construction and performance can be found in Ref.~\cite{Azzolini:2018tum_C0det} and references therein.

\subsection{Data production}
This work is based on data collected with \cuz between June 2017 and December 2018, for a total exposure of \exposure~(\znsenr).
Two of the enriched crystals, not properly working, and the two with natural Se are not included in this analysis.

The collected data are processed offline using a C++ based analysis framework originally developed by the \qz and CUORE collaborations~\cite{Alduino:2016zrl_Q0Analysis}.
The specific analysis tools developed for scintillating calorimeters in the framework of \cuz are presented in Ref.~\cite{Azzolini:2018yye_C0Analysis,Azzolini:2018oph_C0ExcitedStates}.

The aim of the data production sequence is to extract from each  triggered waveform the corresponding energy release and interaction time.
To improve the signal-to-noise ratio, the data are filtered with a software matched-filter algorithm \cite{Gatti:1986cw,Radeka:1966}.
The filtered amplitude is then corrected for gain instabilities using the reference pulses periodically injected through Si heaters~\cite{Carniti:2017zkr}. 
The corrected amplitude is converted into energy by fitting a parabolic function with zero intercept to the energy of the most intense peaks produced by a $^{232}$Th source periodically positioned close to the cryostat external shield \cite{Azzolini:2018yye_C0Analysis}. The heat released by $\alpha$ and $\beta/\gamma$ of the same energy is slightly different because of the different energy spent in the light channel. To re-calibrate the $\alpha$ events, we identify the most intense $\alpha$ peaks produced by $^{238}$U and $^{232}$Th internal contaminations (see Fig.~\ref{fig:ExpSpectra}), and convert the amplitude to energy using a parabolic function. Data acquired between two calibrations are grouped in a \textit{DataSet} and processed together through the analysis chain.

We compute time coincidences between detectors within a $\pm$20~ms window, optimized by studying the time distribution of physical coincident events collected during calibrations. 
Time coincident events are organized in a multiplet structure, which includes information about the triggered crystals and the total energy released in the detector.
Since the total event rate is approximately 50~mHz, the probability of accidental (i.e. causally unrelated) coincidences is relatively small ($\sim 10^{-3}$). 

Finally, we implement a series of event selection cuts in order to maximize our sensitivity to physics events~\cite{Azzolini:2018yye_C0Analysis}. 
Periods of cryostat instability and malfunction are excluded on a crystal-by-crystal basis. 
A time veto around each event (\unit[4]{s} before and \unit[4]{s} after) is applied to remove piled up events.
We exploit heater pulses to calculate the trigger efficiency (i.e. the probability that an event is detected and reconstructed at the right energy) and the pile-up cut efficiency \cite{Alduino:2016zrl_Q0Analysis}.
We select particle events by requiring a non-zero light signal simultaneously recorded by Ge-LDs. The efficiency of this cut is evaluated by analyzing time coincident events in two crystals, providing a pure sample of particle events, given the negligible probability of random coincidences~\cite{Alduino:2016vtd_Q02nbb}.
The combined efficiency has a constant value of $\varepsilon_{C}=$(95.7 $\pm$ 0.5) \% above 150~keV.

\subsection{Tagging of $\alpha$ particles}

\begin{figure}[t!]
\begin{center}
\includegraphics[width=.45\textwidth]{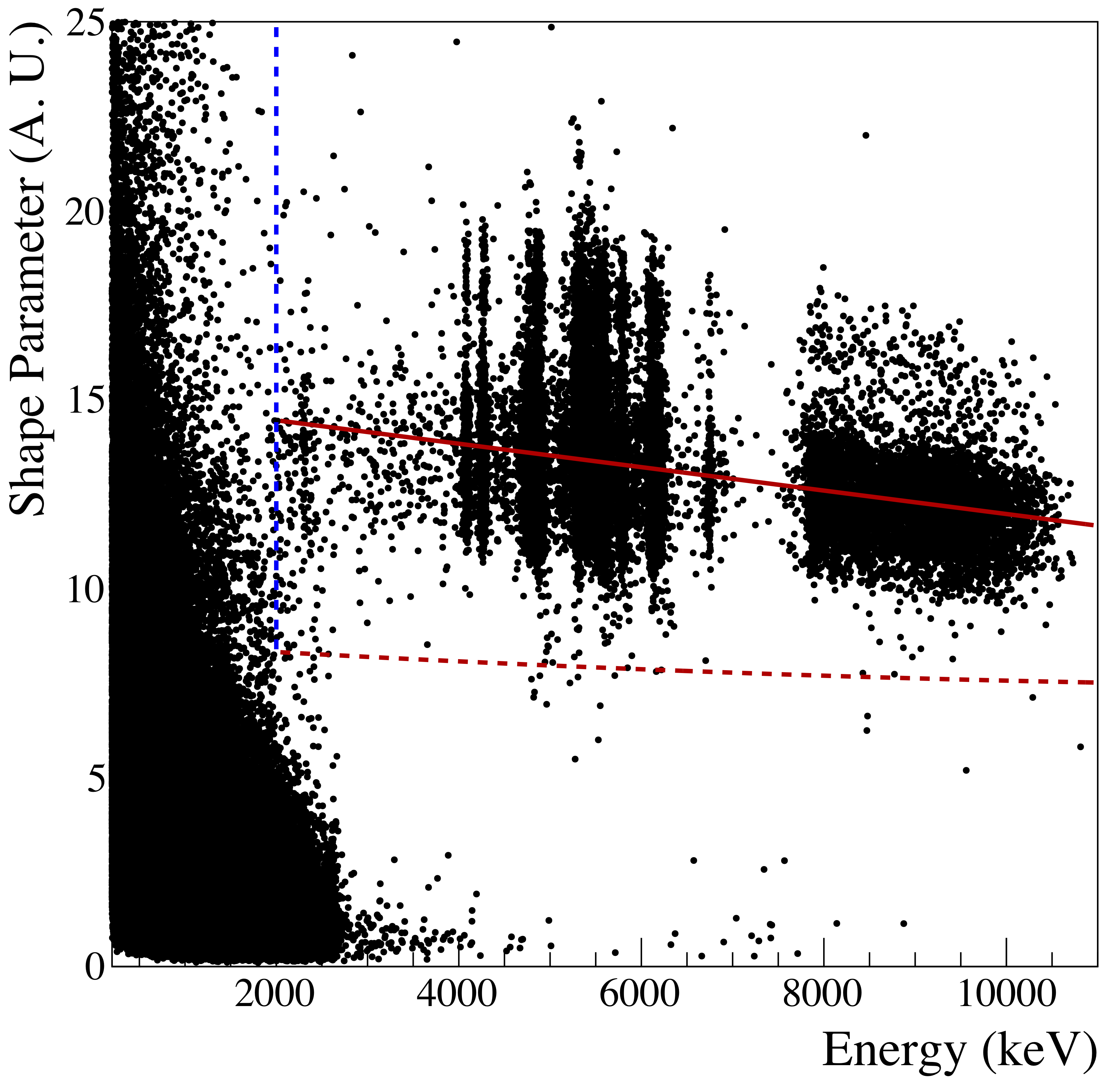}
\caption{Shape Parameter (SP) of light pulses as a function of particle energy. The red solid line is the mean value of $\alpha$ particles SP, while the red dashed line at $\mu_{\alpha}(E)-3\times\sigma_{\alpha}(E)$ is the boundary used to discriminate $\alpha$ from $\beta/\gamma$ events. The blue dashed line at 2~MeV shows the energy below which the particle identification is not applied.} 
\label{fig:alphaID}
\end{center}
\end{figure}

The events generated by $\alpha$ particle interaction are identified relying on the different time development of their light pulses with respect to the ones produced by $\beta$/$\gamma$ interactions~\cite{Artusa:2016maw}. Such different pulse shape is quantified by the Shape Parameter (SP):
\begin{equation}
 \text{SP} = \frac{1}{Aw_r}\sqrt{\sum_{i=i_M}^{i_M+\omega_r}\left(y_i-A \cdot S_i\right)^2}\label{tvr}  
\end{equation}
where $y_i$ are the samples of the filtered light pulse, $A$ is its maximum amplitude, and $S_i$ are the samples of the filtered average pulse scaled to unitary amplitude and aligned to $y_i$. 
The summation starts from the index i$_M$ corresponding to the position of the maximum and runs for w$_r$ points ($\sim$ 50) corresponding to the right width at half maximum of S$_i$. 
The average pulse is made selecting only $\beta/\gamma$ events in the energy range 1.8$-$2.64~MeV of $^{232}$Th calibration measurements, with the method described in Ref.~\cite{Azzolini:2018yye_C0Analysis}.
As a consequence, the SP of $\alpha$ events is much higher than the SP of $\gamma$ events.
Fig.~\ref{fig:alphaID} shows the values of SP as a function of energy in CUPID-0 data. Particle identification is difficult below 2~MeV and, thus, it is exploited only above this energy.

To discriminate $\alpha$ from $\beta/\gamma$ events, we calculate the mean and the standard deviation of the $\alpha$ particle SP as a function of energy and we set a boundary at $\text{SP}=\mu_{\alpha}(E)-3\times\sigma_{\alpha}(E)$. 
The $\mu_{\alpha}(E)$ and $\sigma_{\alpha}(E)$ values of SP are calculated excluding the $\beta/\gamma$ events with SP$<6$ (i.e. the cut used in Ref.~\cite{Azzolini:2018dyb_C0PRL} to select the $\beta/\gamma$ events). 
The discrimination boundary adopted in this analysis allows to correctly identify energy depositions by $\alpha$ particles with a probability of 99.9\% at all energies greater than $2$~MeV. This boundary is higher than the $\mu_{\beta/\gamma}(E)+5\times\sigma_{\beta/\gamma}(E)$ contour of $\beta/\gamma$ SP distribution, thus we select $\beta/\gamma$ events with unitary efficiency.
The expected number of wrongly identified $\alpha$ events introduces a negligible contamination in the $\beta/\gamma$ spectrum up to $\sim5$~MeV.

\section{Background Analysis}
\label{sec:BkgAnalysis}
Based on the results from previous experiments~\cite{Alduino:2016vtd_Q02nbb,Alduino:2017qet_BkgBudget,Arnold:2018tmo}, the sources of background expected in \cuz are:
\begin{itemize}
\item the \bbd decay of \seod;
\item contaminations of the experimental setup (including the detector itself, the cryostat and the shielding) due to ubiquitous natural radioisotopes of \th, \u and \udtc decay chains, and \kq;
\item isotopes produced by cosmogenic activation of detector materials, such as \cosz and $^{54}$Mn in copper and \znsc in ZnSe;
\item cosmic muons, environmental $\gamma$-rays and neutrons.
\end{itemize}

For a better disentanglement of background sources, we exploit the discrimination of $\alpha$ versus $\beta/\gamma$ events and the detector modular design, that allows to tag events producing simultaneous energy depositions in different ZnSe crystals. We then build the following energy spectra:
\begin{itemize}
\item \monebeta is the energy spectrum of $\beta/\gamma$ events that triggered only one bolometer (multiplicity 1 events); this spectrum includes also $\alpha$ events with E$<$2~MeV that, however, provide a minor contribution; 
\item \monealfa is the energy spectrum of multiplicity 1 events produced by $\alpha$ particle interactions;
\item \mtwo is the energy spectrum of events that simultaneously triggered two bolometers (multiplicity 2 events), built with the energies detected in each crystal;
\item \mtwosum is the energy spectrum associated to multiplicity 2 events that contains, for each couple of time coincident events, the total energy released in both crystals. 
\end{itemize}
Events with higher order multiplicity are used to evaluate the contribution of muons, that generate electromagnetic showers triggering several bolometers at the same time. 

\begin{table}[t!]
\begin{center}
\caption{$\gamma$-lines identified in the CUPID-0 data and their counting rate in \monebeta and \mtwosum spectra.}
\begin{tabular}{rcrr}

\hline
Energy 	& Isotope	& Rate \monebeta	&	Rate \mtwosum   \\
(keV)	& 			& {(counts/(\kgyr))}   &	{(counts/(\kgyr))} \\
\hline
835		& \mncq		& 46  $\pm$ 11	&   20  $\pm$ 3 \\
911		& \acddo	& 57  $\pm$ 10	&   15  $\pm$ 3 \\
969		& \acddo	& 39  $\pm$ 10	&   14  $\pm$ 3 \\
1116	& \znsc		& 639  $\pm$ 17	&   234  $\pm$ 6 \\
1173	& \cosz		& 35  $\pm$ 9	&   7  $\pm$ 4 \\
1332	& \cosz		& 28  $\pm$ 8	&   11  $\pm$ 2 \\
1461	& \kq		& 230  $\pm$ 12	&   100  $\pm$ 4 \\
1765	& \biduq	& 15  $\pm$ 6	&   5.7  $\pm$ 1.4 \\
2615	& \tld		& 45  $\pm$ 3	&   23.1  $\pm$ 1.7 \\
\hline
\end{tabular}
\label{tab:GammaLines}
\end{center}
\end{table}

\begin{table}[t!]
\begin{center}
\caption{Counting rate of $\alpha$-peaks due to \u, $^{235}$U and \th decay chains. All the isotopes appear only as Q-value peak (Q), except $^{210}$Po that shows also the $\alpha$-peak. Since the line shape is not perfectly Gaussian and some peaks are partially overlapped, the activity evaluation is affected by a systematic error up to 10\%, depending on the method used to evaluate the net number of counts in the $\alpha$ peaks. A few lines are depleted by pile-up effects, or summed to $\beta$ coincident events, therefore their activities cannot be directly evaluated.}
\begin{tabular}{cclr}

\hline
Isotope 				& 	Energy  &    & Rate \monealfa	\\
						& 	(MeV)	&    & {(counts/(\kgyr))} 	\\
\hline
$^{232}$Th 				& 	4.08	& Q  & 	80  $\pm$ 3 	\\
$^{228}$Th 				& 	5.52	& Q  & 	384  $\pm$ 7	\\
$^{224}$Ra 				& 	5.79	& Q  & 	332  $\pm$ 7	\\
$^{220}$Rn + $^{216}$Po	& 	--		& -- & 	pile up: $\alpha$ + $\alpha$	\\
$^{212}$Bi (B.R. 36\%)	& 	6.21	& Q  & 	140  $\pm$ 4	\\
$^{212}$Bi + $^{212}$Po	& 	E$_{\beta}$ + 7.83 & --  & 	pile up: $\beta$ + $\alpha$	\\
\hline
$^{238}$U 				& 	4.27	& Q  & 	160  $\pm$ 4	\\
$^{234}$U + $^{226}$Ra 	&$\sim$4.87	& Q  & 	716  $\pm$ 11	\\
$^{230}$Th 				& 	4.77	& Q  & 	161  $\pm$ 4	\\
$^{222}$Rn 				& 	5.59	& Q  & 	531  $\pm$ 8	\\
$^{218}$Po 				& 	6.12	& Q  & 	536  $\pm$ 7	\\
$^{214}$Bi + $^{214}$Po	& 	E$_{\beta}$ + 8.95 & --  & 	pile up: $\beta$ + $\alpha$	\\
$^{210}$Po 				& 	5.41	& Q  		& 	174  $\pm$ 5	\\
$^{210}$Po 				& 	5.30	& $\alpha$  & 	392  $\pm$ 10	\\
\hline
$^{231}$Pa 				& 	5.15	& Q  & 	 8.8  $\pm$ 1.3		\\
$^{211}$Bi 				& 	6.75	& Q  & 	 14.3  $\pm$ 1.4	\\
\hline
$^{147}$Sm 				& 	2.31	& Q  & 	 4.2  $\pm$ 0.7		\\
\hline
\end{tabular}
\label{tab:AlfaLines}
\end{center}
\end{table}

\begin{figure*}[t!]
\begin{center}
\includegraphics[width=.49\textwidth]{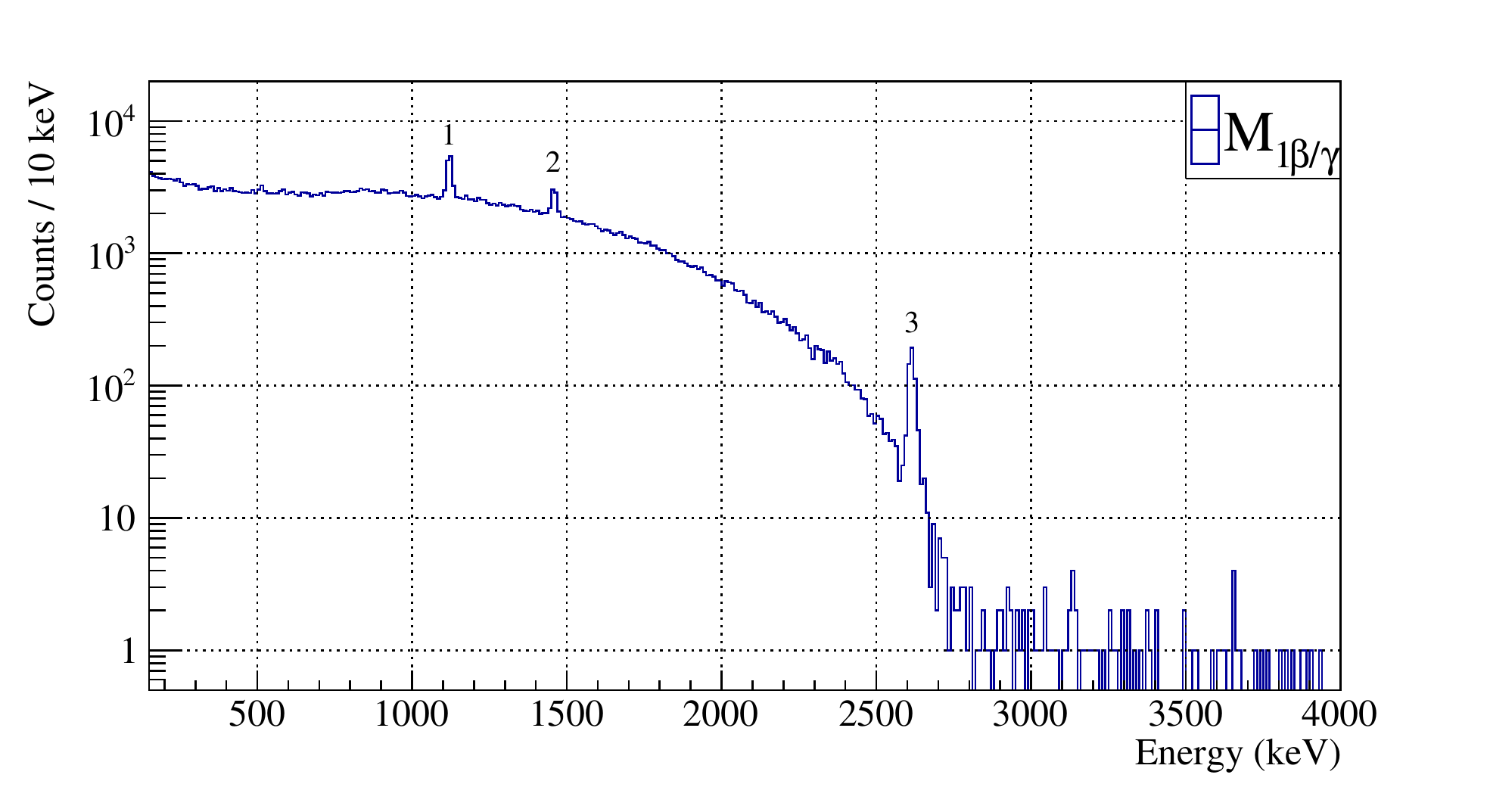}
\includegraphics[width=.49\textwidth]{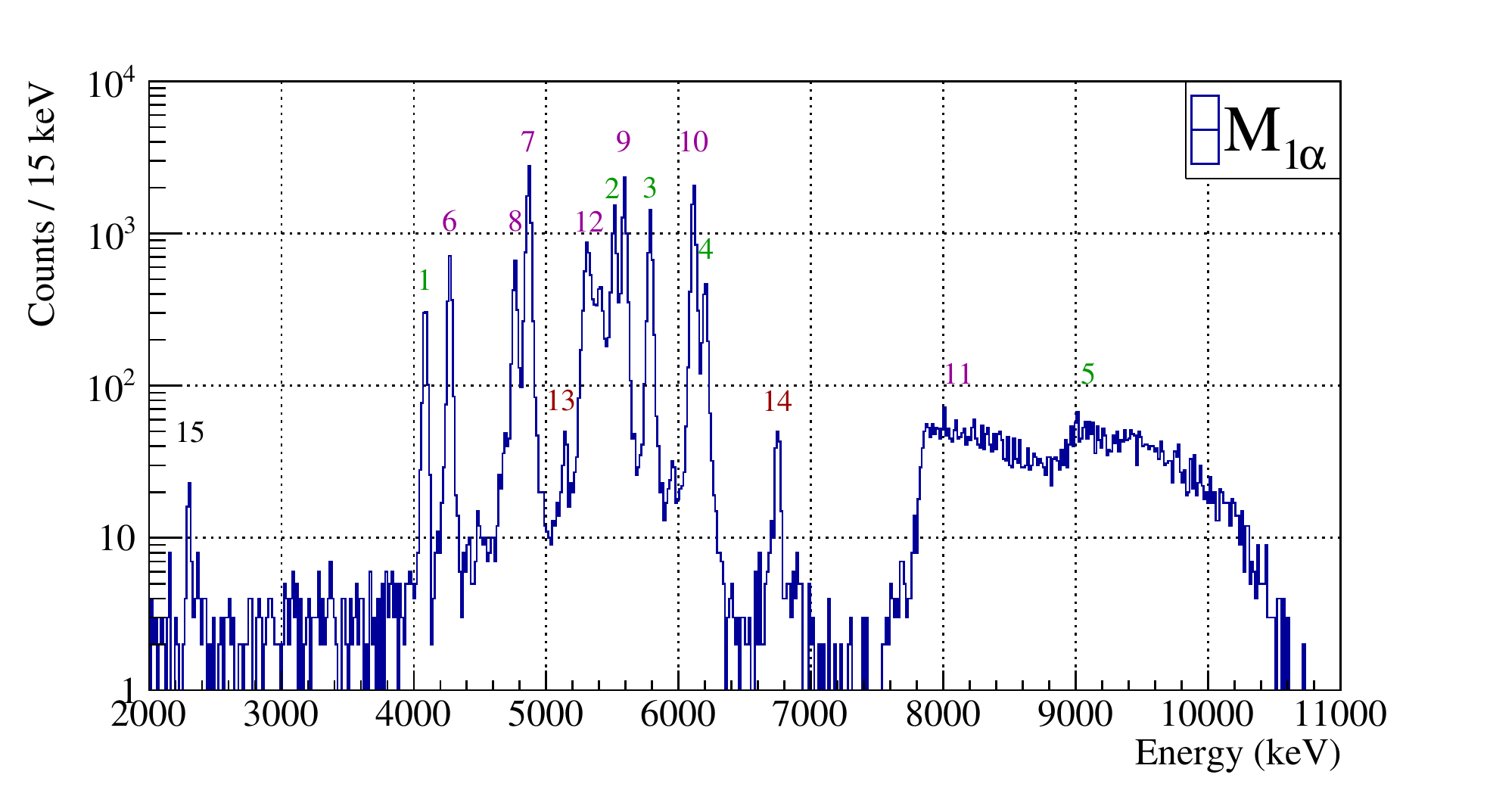}
\includegraphics[width=.49\textwidth]{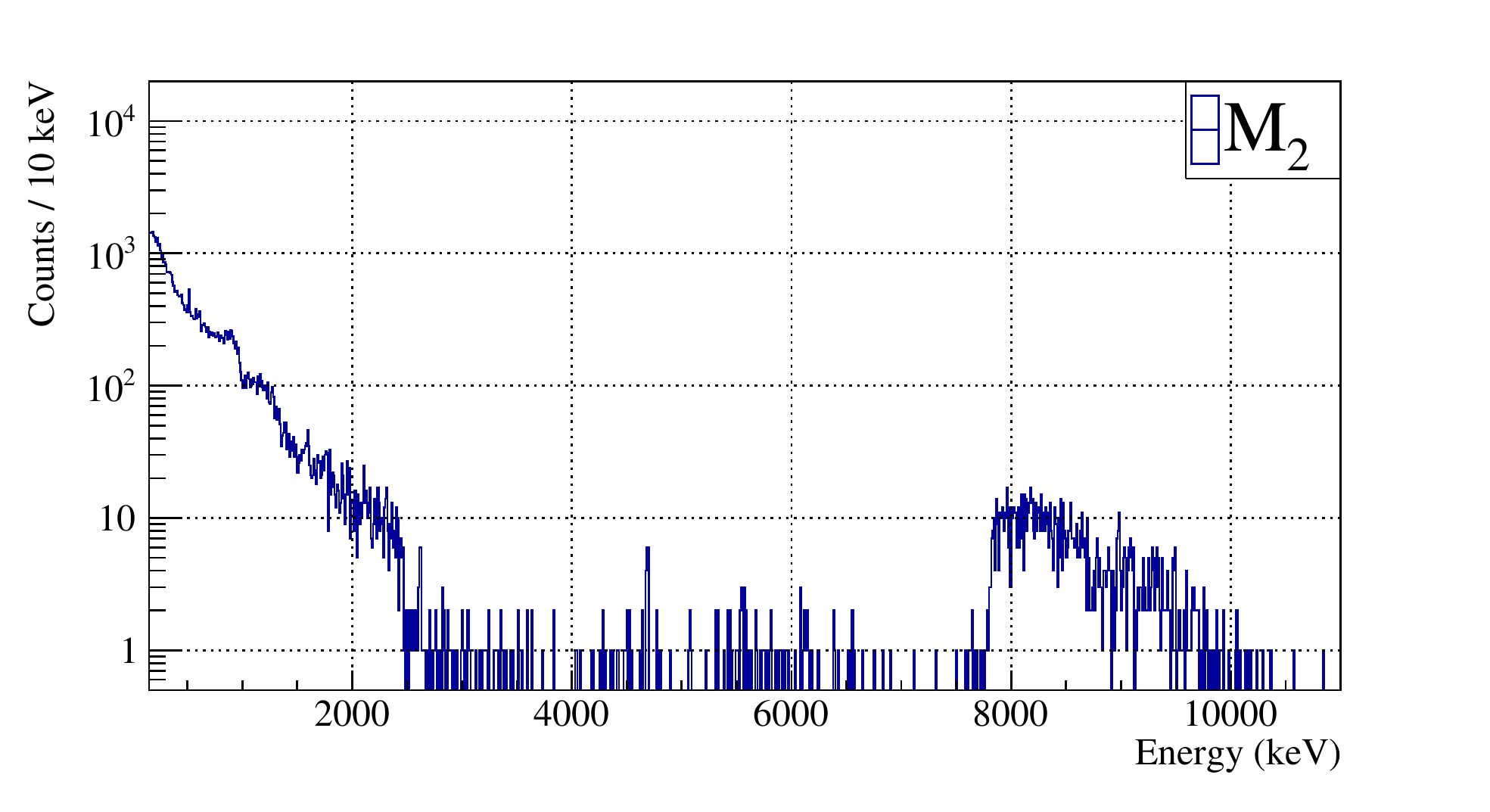}
\includegraphics[width=.49\textwidth]{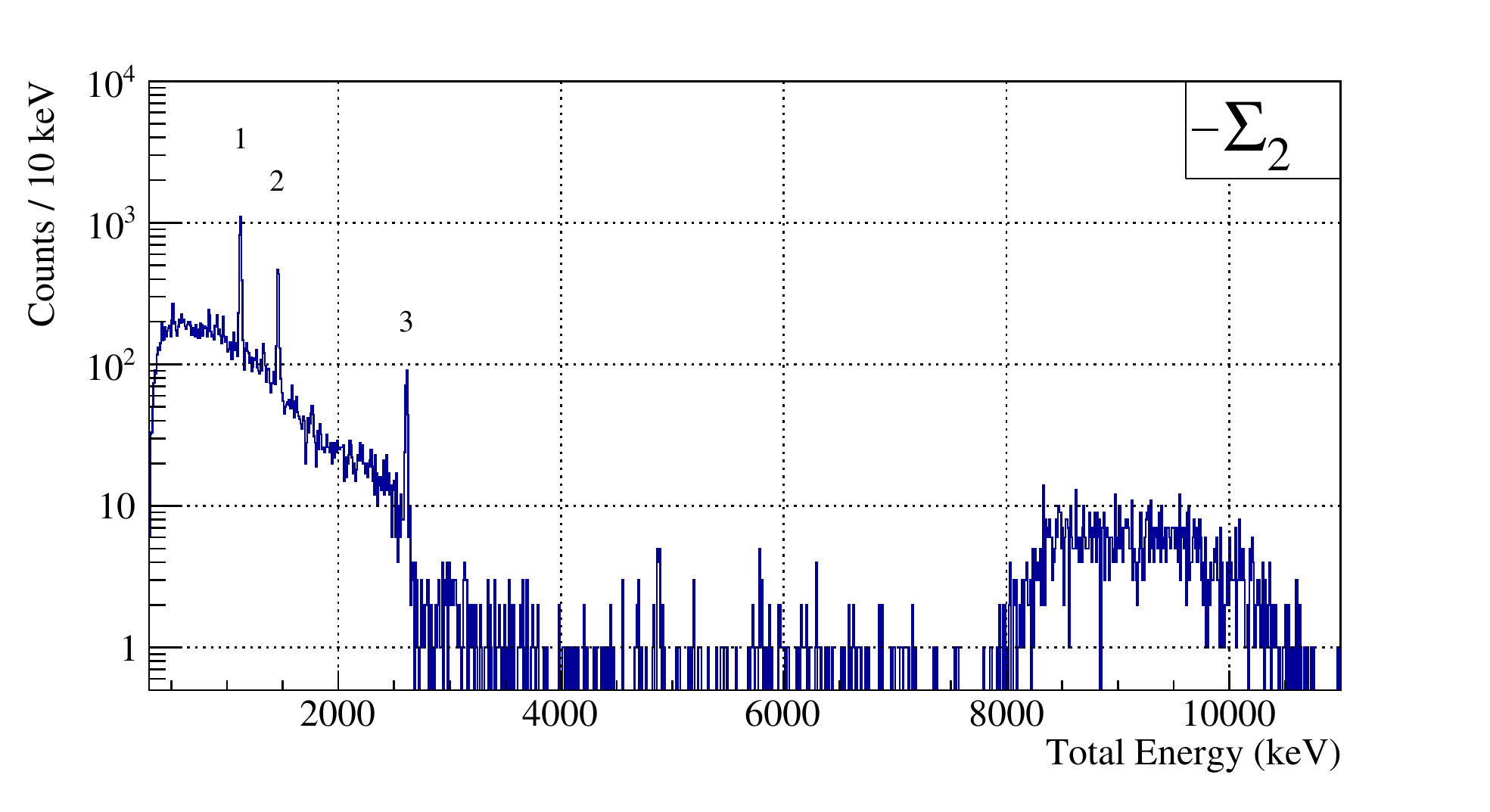}
\caption{\textit{Top left}: \cuz \monebeta spectrum with the following peak labeling: (1) $^{65}$Zn, (2) $^{40}$K, (3) $^{208}$Tl. \textit{Top right}: \monealfa spectrum with the following peak labeling: (1) $^{232}$Th, (2) $^{228}$Th, (3) $^{224}$Ra, (4) $^{212}$Bi, (5) $^{212}$Bi + $^{212}$Po, (6) $^{238}$U, (7) $^{234}$U + $^{226}$Ra, (8) $^{230}$Th, (9) $^{222}$Rn, (10) $^{218}$Po, (11) $^{214}$Bi + $^{214}$Po, (12) $^{210}$Po, (13) $^{231}$Pa, (14) $^{211}$Bi, (15) $^{147}$Sm. \textit{Bottom left}: \mtwo spectrum. \textit{Bottom right}: \mtwosum spectrum with the same labels used for \monebeta peaks.}
\label{fig:ExpSpectra}
\end{center}
\end{figure*}

In Figure~\ref{fig:ExpSpectra} we show the \monebeta, \monealfa, \mtwo and \mtwosum experimental spectra, with labels on the signatures used to identify the background sources. 

The main component of the \monebeta spectrum is the continuum produced by the \bbd decay of \seod. The most intense $\gamma$ lines exceeding this continuum are produced by the decays of \znsc, \kq, and \tld and are clearly visible also in the \mtwosum spectrum. In Table~\ref{tab:GammaLines} we report the counting rates of these $\gamma$ lines and of other smaller peaks attributable to the decays of \cosz, \mncq, \acddo, and \biduq. 

The peaks observed in the \monealfa spectrum are due to $\alpha$ decays occurring in ZnSe crystals or in the detector components directly facing them. These peaks are produced by isotopes belonging to \th, \u and \udtc decay chains, and by $^{147}$Sm, a natural long lived isotope with half-life equal to 1.06$\times$10$^{11}$~yr~\cite{Sm147}.

In Table~\ref{tab:AlfaLines}, we report the counting rates and the energies of the $\alpha$ peaks. 
All the main peaks in the \monealfa spectrum (except the 5.3~MeV line of \podd) are centered at the Q-value of the $\alpha$ decays. This means that the corresponding radioisotopes are located in the bulk or near the surface of ZnSe crystals, because the energy of both $\alpha$ and nuclear recoil is detected.
The counts in the peaks have been evaluated by means of Gaussian fits with linear background subtraction. Taking into account that the line shape is not perfectly Gaussian and that some peaks are partially overlapped, the rates in Table~\ref{tab:AlfaLines} are affected by systematic error up to $\sim$10\%. 
By analyzing the counting rates of the isotopes in each decay chain, we identify the breaking points of secular equilibrium. 
In \th chain, the progenitor \th has a lower rate with respect to the daughters isotopes, thus we can infer a breaking point at $^{228}$Th (that can also be at $^{228}$Ra, since there is no signature for this isotope). 
In \u chain, data interpretation is complicated by the fact that the peak at $\sim$4.87~MeV includes counts from both $^{234}$U and $^{226}$Ra. However, the $^{226}$Ra activity is constrained to be the same of its daughters $^{222}$Rn and $^{218}$Po, which have relatively short half-life. Therefore, we can infer that the first part of the chain from \u to $^{230}$Th is at equilibrium and that there are two breaking points: the first at $^{226}$Ra and the second at $^{210}$Pb.

At energies greater than 7.8~MeV, we observe a continuum spectrum with a double bump shape. The first bump is produced by the $^{214}$Bi--$^{214}$Po decay sequence in \u chain, while the second, starting at 8.9~MeV, is due to the $^{212}$Bi--$^{212}$Po decay sequence of \th chain. In both Bi--Po sequences, the Po half-life is much shorter than the characteristic rise time of thermal pulses (few ms) produced in bolometers, therefore the energy released by Po $\alpha$ decays sums up with the energy deposited by Bi $\beta$ decays. In data production, these events are tagged and calibrated as $\alpha$, because most of the energy is released by the $\alpha$ particle interaction. As a consequence, the energy of $\beta$ component of these events is underestimated by approximately 20\%, due to the different calibration of $\alpha$ and $\beta/\gamma$ energy depositions. 
The Bi--Po signature is present also in the \mtwo and \mtwosum spectra, because  $\beta$ particles and the associated $\gamma$ can cross the reflecting foils around ZnSe crystals and produce \mtwo events. 
On the contrary, since the range of $\alpha$ particles is lower than reflecting foil thickness, the \mtwo and \mtwosum spectra have a small number of events in the range of $\alpha$-lines from 4 to 7~MeV.

\section{Monte Carlo simulations}
\label{sec:MC}
\begin{figure*}[!htb]
\begin{center}
\includegraphics[width=.80\textwidth]{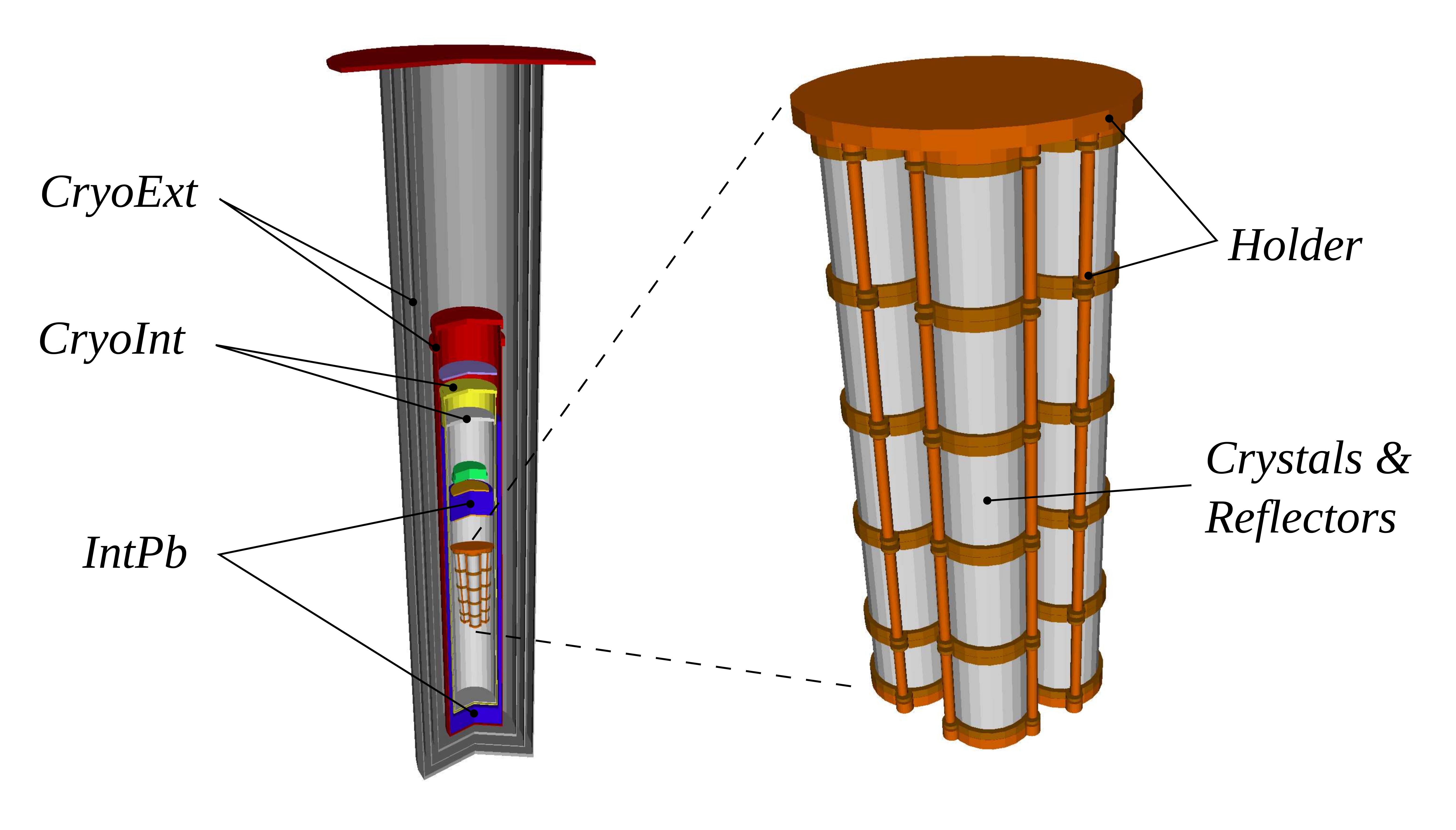}
\caption{Graphic view of the CUPID-0 experimental setup as modeled in \textsc{Geant4} with the {\it Arby} toolkit. External lead and neutron shield are included in the MC but not represented here.}
\label{fig:Arby}
\end{center}
\end{figure*}

The background sources identified through data analysis are simulated with a Monte Carlo toolkit, called \textit{Arby}, based on the \textsc{Geant4} toolkit~\cite{Geant4}, version 4.10.02. 
The radioactive decays from the various background sources can be generated in any volume or surface of the \cuz detector, cryostat and shielding implemented in \textit{Arby}.
The primary and any secondary particles are then propagated through the \cuz geometry using the Livermore physics list. 
The energy deposited in ZnSe crystals is recorded in the Monte Carlo output together with the time at which the interaction occurred. The fraction of energy released by any particle type is also recorded to allow particle identification.
Radioactive decays are implemented using the G4RadioactiveDecay database. The decay chains of \th, \u, and \udtc can be simulated completely or in part, to reproduce breaks of secular equilibrium. The \bbd simulation is generated under the single-state dominance hypothesis (SSD) in the framework of the Interacting Boson Model (IBM)~\cite{Kotila:2012zza}, while the generation of external muons is described in Ref.~\cite{Andreotti:2009dk}.

In order to implement the detector response function and data production features in the Monte Carlo data, we reprocess the \textit{Arby} output with a dedicated code.
In particular, to account for detector time resolution, we sum energy depositions that occur in the same crystal within a $\pm$5~ms window. 
The experimental energy resolution is reproduced by applying a Gaussian smearing function with linearly variable width based on measured FWHM of $\gamma$ and $\alpha$ lines.
The energy threshold of each detector is modeled with an error function that interpolates the experimental data of trigger efficiency versus energy. These data are collected in dedicated runs in which the heater is used to generate pulses with variable amplitudes (then converted into particle equivalent energies) and the efficiency is calculated, for each pulse amplitude, as the ratio between triggered and generated pulses.
Exactly as done in experimental data production, events depositing energy in different crystals within $\pm$20~ms window are combined into multiplets and pile-up events (see Section~\ref{sec:exp}) in the same crystal are discarded.
Finally, exploiting the information about the type of particle depositing energy, we reproduce in the Monte Carlo data the same event selection applied in the experimental data to produce the \monealfa and \monebeta spectra. Particularly, we include in the \monealfa spectrum (with efficiency $>99.9\%$ at E$>2$~MeV) not only the $\alpha$ events, but also the heterogeneous $\beta/\gamma+\alpha$ events due to Bi--Po decay sequences, that in the experimental data are tagged as $\alpha$ events (see Fig.~\ref{fig:alphaID}).

To model the cryostat and its shielding we take as reference the scheme developed for the \qz background model~\cite{Alduino:2016vtd_Q02nbb}, implementing the geometry changes made in \cuz. Concerning particle generation, we group together the components that are made of the same material (and thus share equal contaminant concentration) or that cannot be disentangled as they produce degenerate spectra, given the counting statistics of the experimental data.
In Fig.~\ref{fig:Arby} we show the geometry of \cuz cryostat and detector as implemented in \textit{Arby}. The neutron and modern lead (\pbext) external shields, even if not represented in the figure, are implemented in MC simulations as well (for a detailed scheme and description of these shields see Ref.~\cite{Alduino:2016vtd_Q02nbb}). 
The cryostat components where the background sources are generated are the following:
\begin{itemize}
\item the Cryostat External Shields (\cryoext) include the Inner Vacuum Chamber (IVC), the super-insulation layers, the Outer Vacuum Chamber (OVC), and the main bath, whose spectra are degenerate;
\item the Cryostat Internal Shields (\cryoint) group the 600~mK and the 50~mK shields, that are made of the same copper;
\item the Internal Lead Shield (\pbint) is inserted between the IVC and the 600~mK shield and is made of low background ancient Roman lead. 
\end{itemize}

The \cuz detector itself, reconstructed with high detail in MC simulations, is made of three main components where the background sources are generated:
\begin{itemize}
\item the \holder is the supporting structure for the detectors and is made of a special copper alloy (NOSV copper produced by Aurubis company) suitable for cryogenic use and cleaned according to protocols developed in CUORE~\cite{Alessandria:2012zp};
\item the \crystals are ZnSe cylinders with heights and positions mirroring the real experimental setup;
\item the \reflectors are the foils laterally surrounding the crystals. This component is also used to account for the minor contribution from light detectors (see Tab.~\ref{tab:Limits}), from the amount ($<$ 15 cm$^2$/crystal) of copper surface directly facing the edges of ZnSe crystals, and from the other small parts close to the crystals (PTFE spacers, NTDs, and wires), whose spectra are degenerate with those of reflecting foils.
\end{itemize}

\section{Background model}
\label{sec:BkgModel}
In the construction of a background model, the crucial step is selecting a representative list of sources for fitting the experimental spectra.
The analysis of $\alpha$ and $\gamma$ lines presented in Section~\ref{sec:BkgAnalysis} allows to identify the most relevant sources to be included in the model. Alongside these sources, there are other contaminations not producing prominent signatures and whose location (or even emitting isotope) cannot be determined with certainty. In this case, the use of all possible sources introduces too many degrees of freedom in the fit and produces highly correlated results.
To avoid this drawback, that would mask the precision of the results, we identify the so-called \textit{reference} model, a well balanced set of sources, selected according to the above criteria. We then perform some tests in which the source list is modified to investigate the uncertainties related to the choice of background sources.

In the following subsections, we describe the sources used for background model. We distinguish among \textit{internal/near} and \textit{external} sources. In the first category, $\alpha$ radiation is not shielded and we must model bulk and surface contamination separately, since they are characterized by different signatures and can produce different counting rates in the \bbz ROI. Conversely, bulk and surface contaminations of external sources produce degenerate spectra and cannot be disentangled. 

\subsection{Internal/near sources}

The \textit{internal/near} sources are located in \crystals and in detector components directly facing them, modeled by \reflectors.
Natural $\alpha$ radiation cannot cross the thickness of reflecting foils and light detectors surrounding the crystals.
As a consequence, the \monealfa spectrum is made up of events from internal/near sources only. 

The breaks of secular equilibrium identified in Section~\ref{sec:BkgAnalysis} are modeled by producing separate Monte Carlo simulations for the different parts of the decay chains, to leave them free to converge on different normalizations when performing the fit.

\begin{figure*}[t!]
\begin{center}
\includegraphics[width=.9\textwidth]{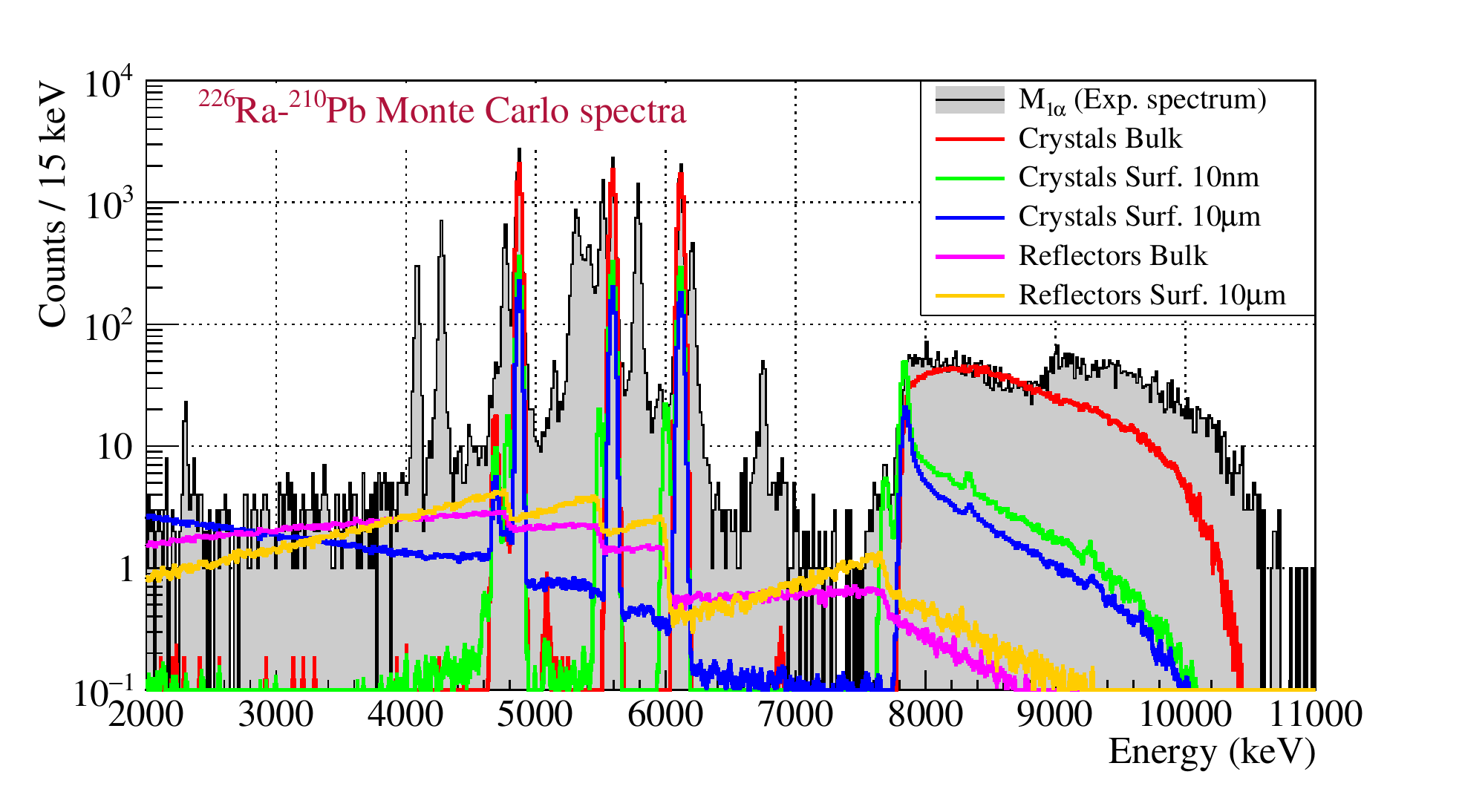}
\caption{Comparison among the spectra of a $^{226}$Ra source simulated in {\it Crystals} or in {\it Reflectors}, both in the bulk and on the surface of each element. Distinctive Q-value peaks characterize all crystal spectra and, in case of surface simulations, there are also smaller peaks at $\alpha$ energies or a continuum spectrum of degraded $\alpha$s, depending on contaminant depth.
Contaminants uniformly distributed in the 70~$\mu$m thickness of \reflectors, or simulated with 10~$\mu$m depth parameter, give rise to similar step-like shaped spectra due to energy degraded $\alpha$s. Normalization of MC spectra is chosen to allow easy comparison with the experimental data. }
\label{fig:MC226Ra}
\end{center}
\end{figure*}

In Fig.~\ref{fig:MC226Ra} we provide an insight of some \monealfa spectra obtained by simulating the decay sequence from $^{226}$Ra to $^{210}$Pb in different positions of \crystals and \reflectors. Bulk simulations are obtained by randomly generating the decays inside a volume. Surface contaminations are simulated with exponential density profiles $e^{-x/\lambda}$ (where $\lambda$ is a changeable depth parameter), used to model not perfectly smooth surfaces and diffusion processes of contaminants.

In the background model, we use three types of simulations for modeling the \monealfa events produced by contaminations of \crystals: 
\begin{itemize}
\item \textit{bulk}, characterized by sharp peaks at Q-value of $\alpha$ decays;
\item \textit{very shallow} with $\lambda=10$~nm, that in addition to the prominent Q-value peaks, exhibit smaller peaks at $\alpha$ energies due to nuclear recoil escapes;
\item \textit{deep surface} with $\lambda=10$~$\mu$m (which is approximately the range of natural $\alpha$ particles in ZnSe), that produce the Q-value peaks over a continuum due to degraded $\alpha$ escapes;
\end{itemize} 

Unlike what was possible to do in \qz background model~\cite{Alduino:2016vtd_Q02nbb}, it is not straightforward to disentangle surface versus bulk contaminations of \crystals, because \mtwo events produced by $\alpha$-escapes from crystal surfaces are completely absent in \cuz data due to the interposition of reflecting foils.
To compensate for the lack of this signature, we developed a new technique based on the time analysis of consecutive $\alpha$ decays.
The ratio between the number of \textit{parent} and time-correlated \textit{daughter} events releasing all the decay energy in the same crystal depends on the contaminant location.
Indeed, given a \textit{parent} event at the Q-value, the probability to detect a time-correlated event at the \textit{daughter} Q-value is nearly one in case of bulk contaminations, whereas it is approximately half in case of surface contaminations, because of $\alpha$-escapes from detector surfaces.
%The ratio between the number of \textit{parent} and \textit{daughter} events releasing all the decay energy in the same bolometer depends on the contaminant location. 
%In fact, both bulk and surface contaminations produce events at the Q-value of \textit{parent} decay, but the probability to detect a time correlated event at the \textit{daughter} Q-value is significantly different, because of $\alpha$-escapes from detector surfaces.
In particular, for \u chain, we count the number of \textit{parent} events at 5.59~MeV peak of $^{222}$Rn, followed by 6.12~MeV \textit{daughter} events produced in the same crystal by $^{218}$Po decay within 3$\times$\vita-dim time window (decay scheme in Eq.~\ref{Eq:uchain}). 

\begin{equation} \label{Eq:uchain}
^{222}\text{Rn} \xRightarrow[\text{Q=5.59~MeV}]{\text{\vita-dim=3.82~d}}\,^{218}\text{Po} \xRightarrow[\text{Q=6.12~MeV}]{\text{\vita-dim=3.11~min}}\, ^{214}\text{Pb}
\end{equation}

Similarly, for \th chain, we look for time correlated events generated by $^{224}$Ra (Q=5.79~MeV) and $^{220}$Rn $\alpha$ decays (scheme in Eq.~\ref{Eq:thchain}). In the latter case, we developed a dedicated tool to recover and count the $^{220}$Rn events which are rejected by standard analysis cuts due to pile-up with $^{216}$Po decay (\vita-dim = 0.145~s).

\begin{equation} \label{Eq:thchain}
^{224}\text{Ra} \xRightarrow[\text{Q=5.79~MeV}]{\text{\vita-dim=3.66~d}}\,^{220}\text{Rn} \xRightarrow[\text{Q=6.4~MeV}]{\text{\vita-dim=55.6~s}}\, ^{216}\text{Po}
\end{equation}

By combining the experimental data of these time coincidences with Monte Carlo simulations, we constrain the ratio of surface versus bulk contaminations of the middle ($^{226}$Ra--$^{210}$Pb) and lower ($^{228}$Ra--$^{208}$Pb) parts of \u and \th decay chains, respectively.
The results of this analysis prove that bulk contaminations have higher activity than surface ones. In particular, we obtain that only $\sim$15\% ($\sim$5\%) of the \textit{parent} events at the Q-value of $^{222}$Rn ($^{224}$Ra) are produced by surface contaminations.
In the background model we exploit this information by setting specific \textit{priors} to constrain the activity of  $^{226}$Ra--$^{210}$Pb and $^{228}$Ra--$^{208}$Pb surface contaminations relative to the bulk ones.

The other contamination of \crystals (see Table~\ref{tab:Results} for the complete list) cannot be constrained with this method and, to avoid getting too much correlated results, are modeled as bulk. This choice does not affect the reconstruction of the background at the \bbz ROI.

The simulations we use for modeling the contaminations of \reflectors are:
\begin{itemize}
\item \textit{very shallow} with $\lambda=10$~nm, producing a sharp peak at $\alpha$ energy;
\item \textit{bulk} or \textit{deep surface} with $\lambda=10$~$\mu$m, both characterized by a continuum spectrum due to degraded $\alpha$ particles.
\end{itemize}
As shown in Figure~\ref{fig:ExpSpectra} and Table~\ref{tab:AlfaLines}, the only $\alpha$ line clearly visible in the \monealfa spectrum results from $^{210}$Po decay. Therefore, the only \textit{very shallow} contamination of \reflectors included in the reference model is the $^{210}$Pb one.
Given the thickness (70~$\mu$m) and low density (0.6~g/cm${^3}$) of reflecting foils, \textit{bulk} and 10~$\mu$m surface contaminations produce nearly degenerate spectra. In the reference model we use the 10~$\mu$m surface ones. 
Based on the measured contaminations of reflector foils reported in Table~\ref{tab:Limits}, we do not include the upper part of \u decay chain (whose contribution is negligible), and we set a prior on \th chain using the the upper limit on $^{228}$Ra contamination. Therefore, the only unconstrained \textit{deep surface} contamination of \reflectors is the lower part of \u chain, which is split into $^{226}$Ra--$^{210}$Pb and $^{210}$Pb--$^{206}$Pb, to allow a break of equilibrium.

\subsection{External sources}

The \textit{external} sources are contaminations in the holder, in the cryostat and in the shields. These sources produce events in the \monebeta, \mtwo, and \mtwosum spectra. The $\gamma$ lines that can be used to identify these sources, besides being few, have limited statistics and, thus, cannot be exploited to directly extract information about the position of contaminations.
For this reason, we take as reference the background model of the \qz experiment, that was operated in the same cryostat as \cuz. Given the lower statistics of \cuz data and the higher \bbd rate, we have to apply further approximations with respect to the \qz model. 
Particularly, we cannot disentangle \th, \u and \kq contaminations in \cryoext from the ones in \pbext, because their spectra are degenerate. 
For the same reason, we merge \holder and \cryoint sources. The only exception is $^{54}$Mn, a cosmogenic-activated isotope with \vita-dim=312~d, that is mainly located in the most recently produced copper of the \holder structure.
\cosz external sources are simulated only in copper components: \cryoint and \cryoext. We use the result from the \qz background model to set a prior for \cosz in \cryoext.
Moreover, we do not include \kq contaminations in \pbint shield, since the \qz model sets an upper limit for this source.
Finally, we simulate a $^{210}$Pb source in \pbext, because the bremsstrahlung produced by $^{210}$Bi decay was found to produce a sizable amount of events in \qz experiment and, thus, is expected to provide a contribution also in \cuz.

\begin{figure*}[t!]
\begin{center}
\includegraphics[width=.8\textwidth]{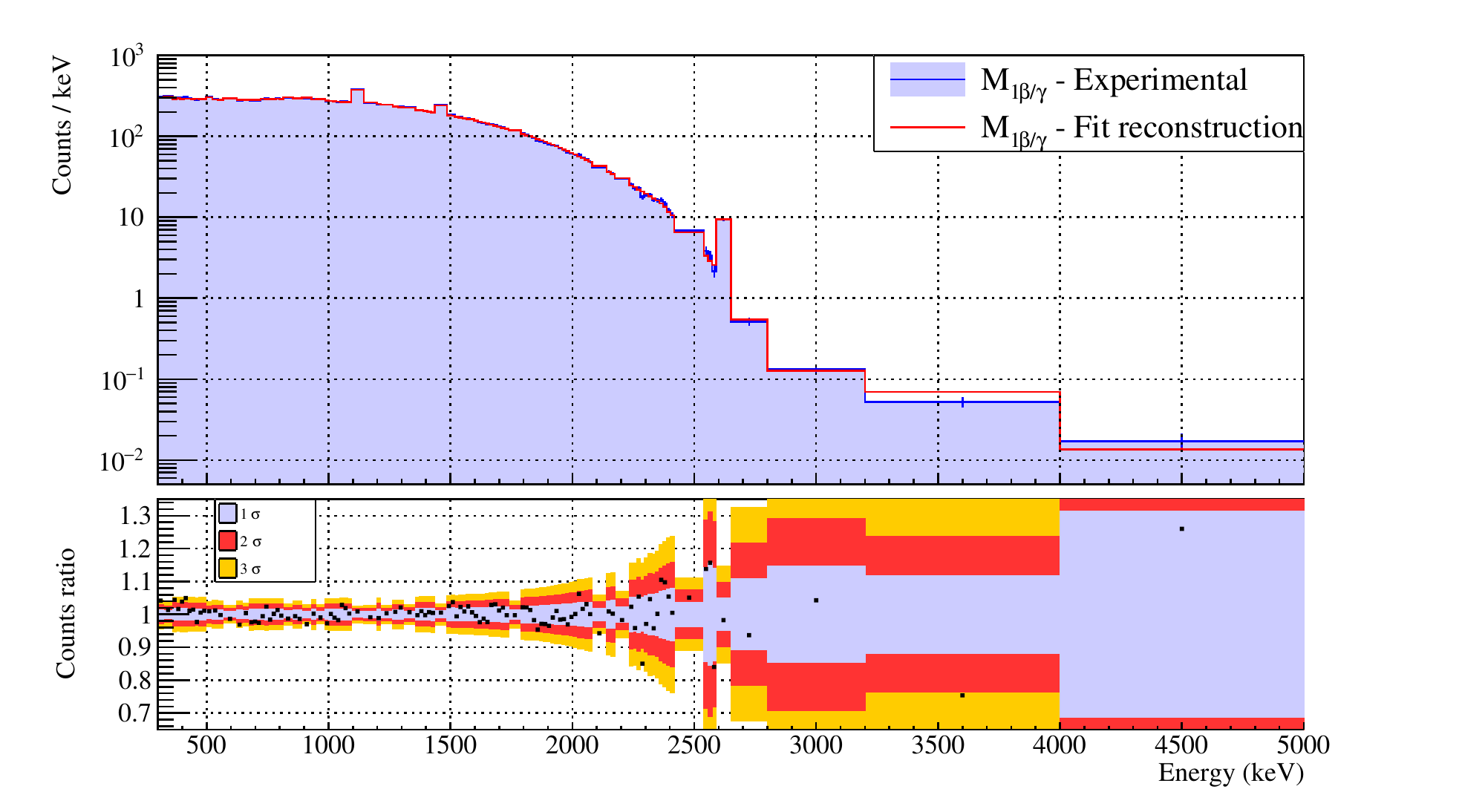}
\includegraphics[width=.8\textwidth]{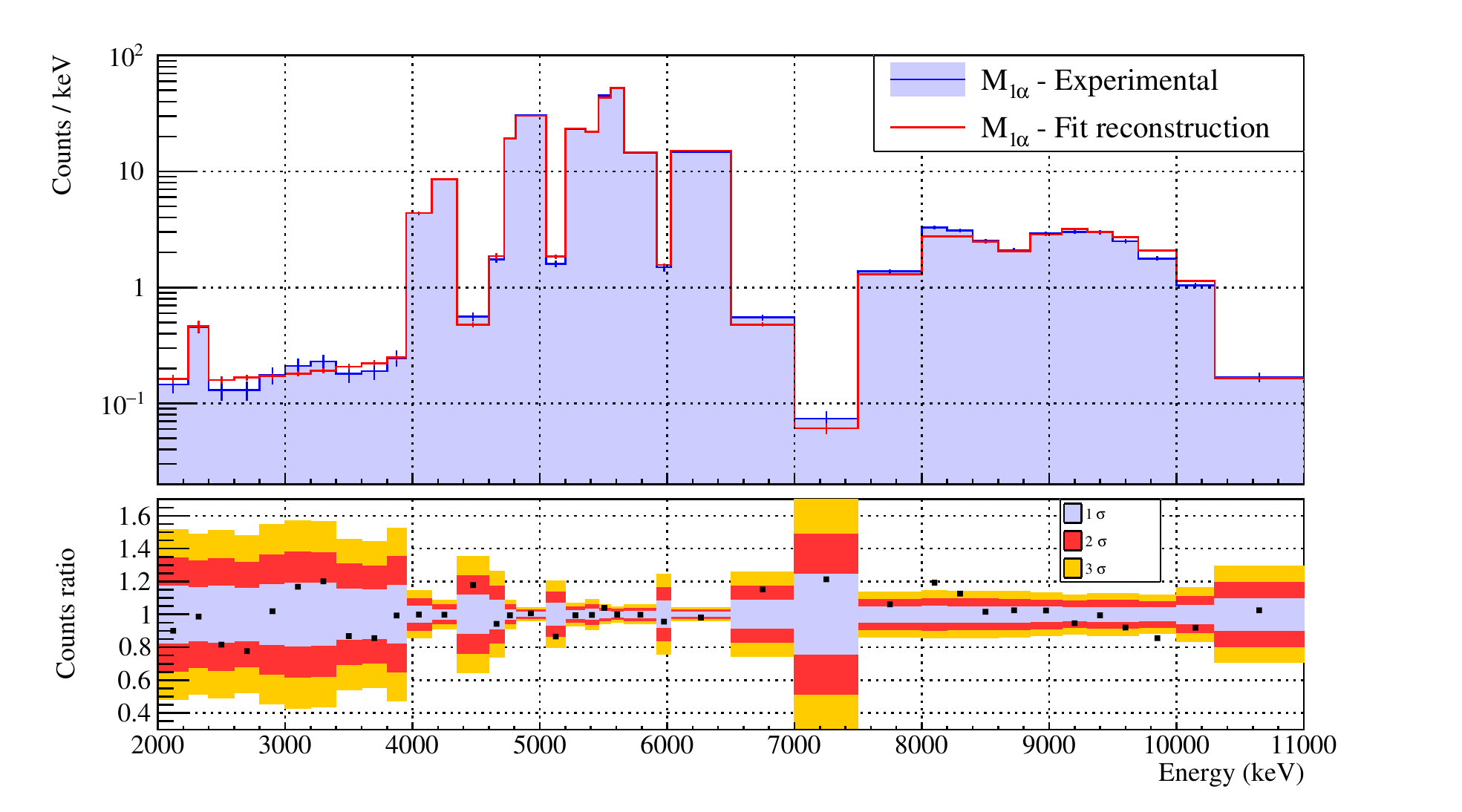}
\caption{
\textit{Top}: comparison between experimental \monebeta spectrum and fit reconstruction. The lower panel shows the bin by bin ratios between counts in the experimental spectrum over counts in the reconstructed one; the corresponding uncertainties at 1, 2, 3 $\sigma$ are shown as colored bands centered at 1.
\textit{Bottom}: same as \textit{Top} for \monealfa spectrum.
}
\label{fig:JAGSM1}
\end{center}
\end{figure*}

\subsection{Environmental sources}

The muon flux, even if strongly suppressed by the Gran Sasso rock overburden, is expected to provide a not negligible contribution to the background in the \bbz ROI. Muons interacting in the detector components can produce several $\gamma$ rays triggering high multiplicity events. We exploit this signature to determine the normalization of the simulated muon spectrum. 
With this method we determine the contribution from muons within a $\pm$15\% systematic uncertainty, depending on the selection of experimental high multiplicity events used to calculate the normalization factor.
The obtained result, which is compatible with measurements performed by other experiments~\cite{Ambrosio:1995cx}, is then used to set a prior for muons in the background model.

The contribution due to environmental neutrons and $\gamma$-rays is negligible, as stated in the \qz background model~\cite{Alduino:2016vtd_Q02nbb}.

\section{Results} % reference fit
\label{sec:Results}

\begin{figure*}[t!]
\begin{center}
\includegraphics[width=.8\textwidth]{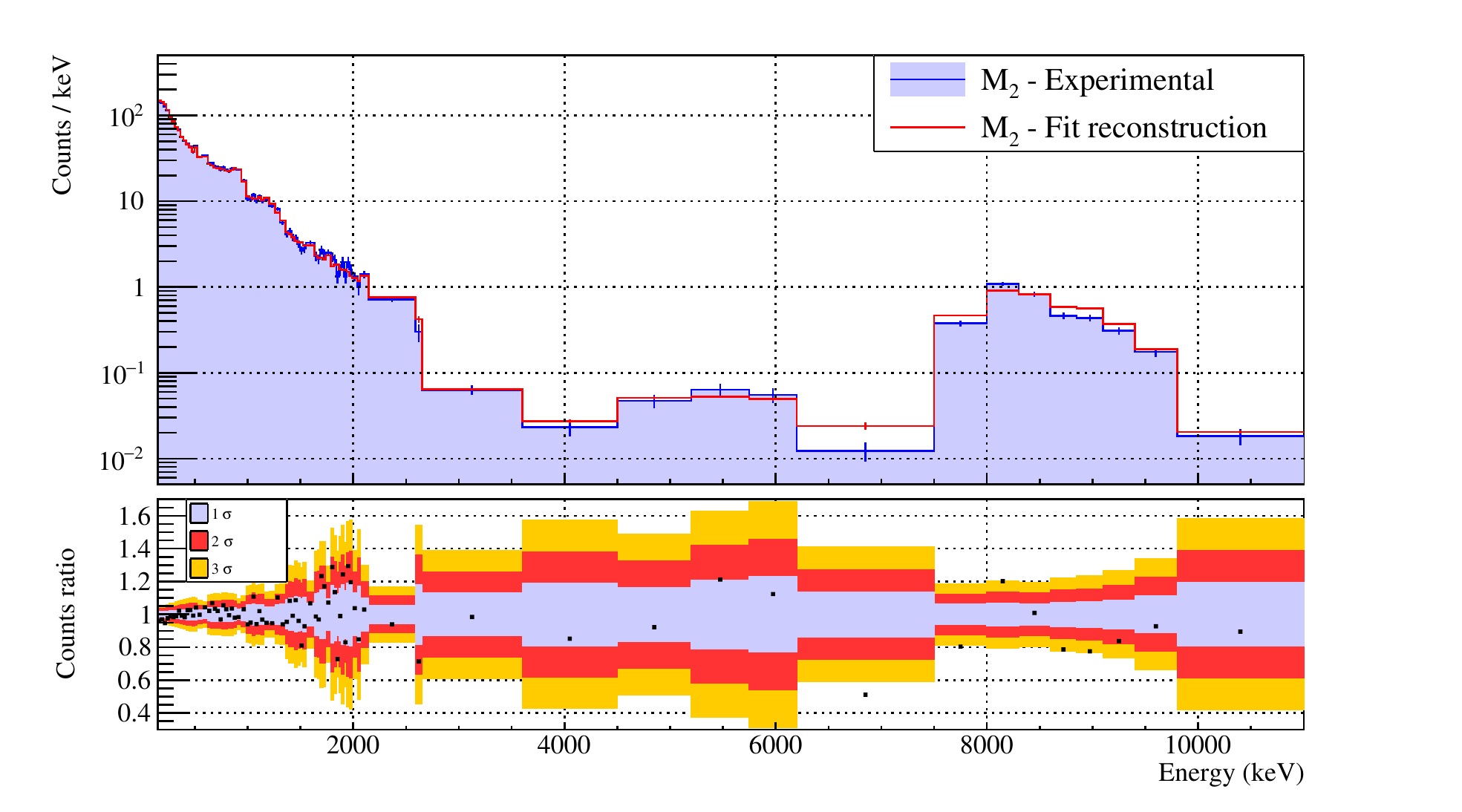}
\includegraphics[width=.8\textwidth]{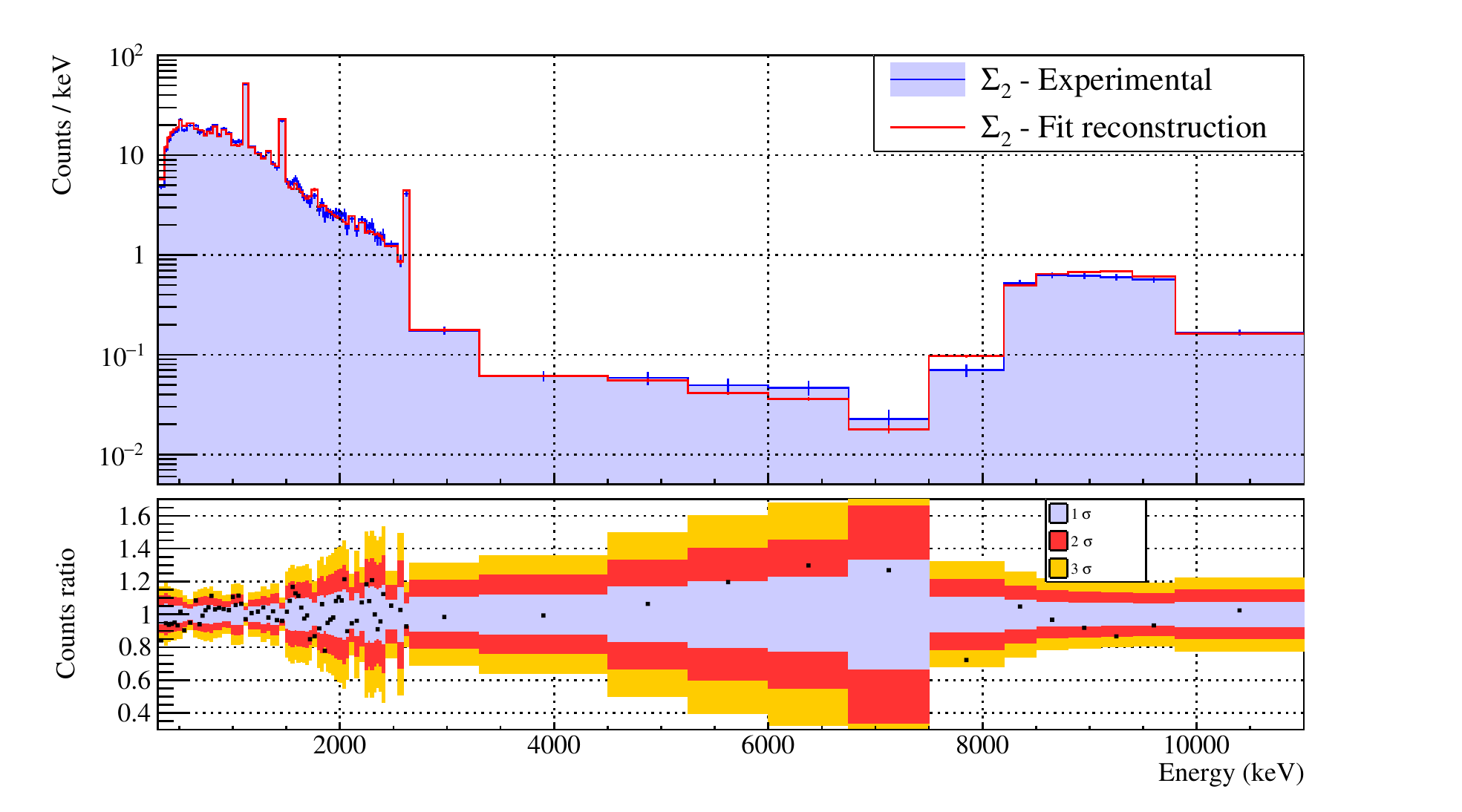}
\caption{\textit{Top}: comparison between experimental \mtwo spectrum and fit reconstruction. The lower panel shows the bin by bin ratios between counts in the experimental spectrum over counts in the reconstructed one; the corresponding uncertainties at 1, 2, 3 $\sigma$ are shown as colored bands centered at 1.
\textit{Bottom}: same as \textit{Top} for \mtwosum spectrum.}
\label{fig:JAGSM2}
\end{center}
\end{figure*}

\begin{table*}[!htb]
\begin{center}
\caption{List of the sources used to fit the CUPID-0 background data in the \textit{reference} model. The columns contain (1) the name of the component where the source is located, (2) the corresponding mass (or surface), (3) the contaminant, (4) the source index used to plot the correlation matrix in Fig~\ref{fig:CorrMatrix}, and (5) the evaluated activity with the statistical uncertainty (limits are at 90\% C.L.). The source activities listed in this table are the result of a model involving some approximations and the corresponding systematic uncertainties are not included here. }
\begin{tabular}{lllll}
\hline
Component &  Mass (kg) & Source & Index & Activity (Bq/kg)\\
\hline

\multirow{12}{*}{\crystals}	&	\multirow{12}{*}{10.5}	&	\bbd	&	1	&	$(9.96\pm0.03)\times10^{-4}$	\\
	&		&	$^{65}$Zn	&	2	&	$(3.52\pm0.06)\times10^{-4}$	\\
	&		&	$^{40}$K	&	3	&	$(8.5\pm0.4)\times10^{-5}$	\\
	&		&	$^{60}$Co	&	4	&	$(1.4\pm0.3)\times10^{-5}$	\\
	&		&	$^{147}$Sm	&	5	&	$(1.6\pm0.3)\times10^{-7}$	\\
	&		&	$^{238}$U--$^{226}$Ra	&	6	&	$(5.51\pm0.10)\times10^{-6}$	\\
	&		&	$^{226}$Ra--$^{210}$Pb	&	7	&	$(1.54\pm0.02)\times10^{-5}$	\\
	&		&	$^{210}$Pb--$^{206}$Pb	&	8	&	$(7.05\pm0.16)\times10^{-6}$	\\
	&		&	$^{232}$Th--$^{228}$Ra	&	9	&	$(2.74\pm0.10)\times10^{-6}$	\\
	&		&	$^{228}$Ra--$^{208}$Pb	&	10	&	$(1.20\pm0.03)\times10^{-5}$	\\
	&		&	$^{235}$U--$^{231}$Pa	&	11	&	$(5.3\pm0.7)\times10^{-7}$	\\
	&		&	$^{231}$Pa--$^{207}$Pb	&	12	&	$(7.8\pm0.4)\times10^{-7}$	\\
\hline									
\holder	&	3.10	&	$^{54}$Mn	&	13	&	$(2.2\pm0.3)\times10^{-4}$	\\
\hline									
\multirow{4}{*}{\cryoint ($^a$) }	&	\multirow{4}{*}{36.9}	&	$^{232}$Th	&	14	&	$<4.5\times10^{-5}$	\\
	&		&	$^{238}$U	&	15	&	$(7\pm3)\times10^{-5}$	\\
	&		&	$^{40}$K	&	16	&	$(3.0\pm0.6)\times10^{-3}$	\\
	&		&	$^{60}$Co	&	17	&	$(6.8\pm1.3)\times10^{-5}$	\\
\hline									
\multirow{2}{*}{\pbint}	&	\multirow{2}{*}{202}	&	$^{232}$Th	&	18	&	$<6.3\times10^{-5}$	\\
	&		&	$^{238}$U	&	19	&	$<7.3\times10^{-5}$	\\
\hline									
\cryoext	&	832	&	$^{60}$Co	&	20	&	$(2.6\pm0.9)\times10^{-5}$	\\
\hline									
\multirow{4}{*}{\pbext ($^b$)}	&	\multirow{4}{*}{24694}	&	$^{232}$Th	&	21	&	$(4.3\pm0.6)\times10^{-4}$	\\
	&		&	$^{238}$U	&	22	&	$(2.5\pm1.2)\times10^{-4}$	\\
	&		&	$^{40}$K	&	23	&	$(2.8\pm0.8)\times10^{-3}$	\\
	&		&	$^{210}$Pb	&	24	&	$7.8\pm0.3$	\\
\hline
Component &  Surface (cm$^2$) & Source & Index & Activity (Bq/cm$^2$)\\
\hline									
\multirow{4}{*}{\crystals}	&	\multirow{4}{*}{2574}	&	$^{226}$Ra--$^{210}$Pb--0.01$\mu$m	&	25	&	$(2.63\pm0.15)\times10^{-8}$	\\
	&		&	$^{228}$Ra--$^{208}$Pb--0.01$\mu$m	&	26	&	$(6.5\pm1.1)\times10^{-9}$	\\
	&		&	$^{226}$Ra--$^{210}$Pb--10$\mu$m	&	27	&	$<2.3\times10^{-9}$	\\
	&		&	$^{228}$Ra--$^{208}$Pb--10$\mu$m	&	28	&	$(4.2\pm1.6)\times10^{-9}$	\\
\hline									
\multirow{4}{*}{\reflectors ($^c$)}	&	\multirow{4}{*}{2100}	&	$^{232}$Th--10$\mu$m	&	29	&	$<7.3\times10^{-10}$	\\
	&		&	$^{226}$Ra--$^{210}$Pb--10$\mu$m	&	30	&	$(8.7\pm1.3)\times10^{-9}$	\\
	&		&	$^{210}$Pb--$^{206}$Pb--10$\mu$m	&	31	&	$(1.0\pm0.5)\times10^{-8}$	\\
	&		&	$^{210}$Pb--$^{206}$Pb--0.01$\mu$m	&	32	&	$(1.43\pm0.02)\times10^{-7}$	\\
\hline									
Muons	&	\multicolumn{2}{l}{  Flux in units of $\mu$/(cm$^{2}$s)		}	&	33	&	 $(3.7\pm0.2)\times10^{-8}$ 	\\
\hline
\noalign{\smallskip}
\multicolumn{5}{l}{($^a$) \cryoint sources include also a minor contribution from \holder bulk contaminations.} \\
\multicolumn{5}{l}{($^b$) \pbext is used to represent also the \cryoext sources, that exhibit degenerate spectra.} \\
\multicolumn{5}{l}{($^c$) \reflectors include also a contribution from light detectors, and from copper surface} \\
\multicolumn{5}{l}{and other parts directly facing the ZnSe crystals.} \\

\end{tabular}
\label{tab:Results}
\end{center}
\end{table*}

\begin{figure}[!htb]
\begin{center}
\includegraphics[width=.49\textwidth]{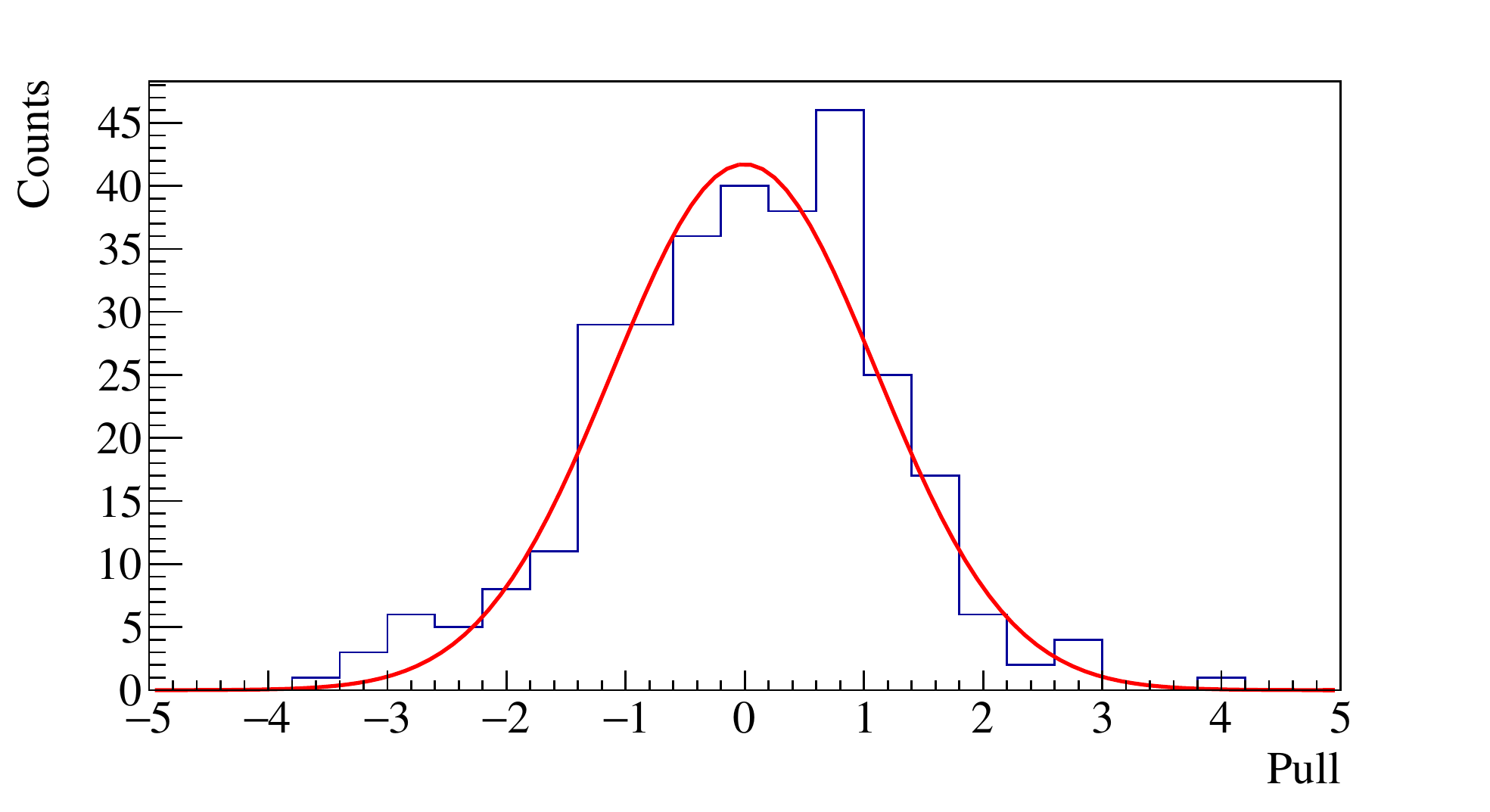}
\caption{Pull distribution obtained from the \textit{reference} fit residuals of all bins. The Gaussian fit to the pull distribution gives $\mu$=0.00$\pm$0.07 and $\sigma$=1.11$\pm$0.06, with $\chi^2$/dof = 18.4/15.}
\label{fig:Pulls}
\end{center}
\end{figure}

\begin{figure}[!htb]
\begin{center}
\includegraphics[width=.49\textwidth]{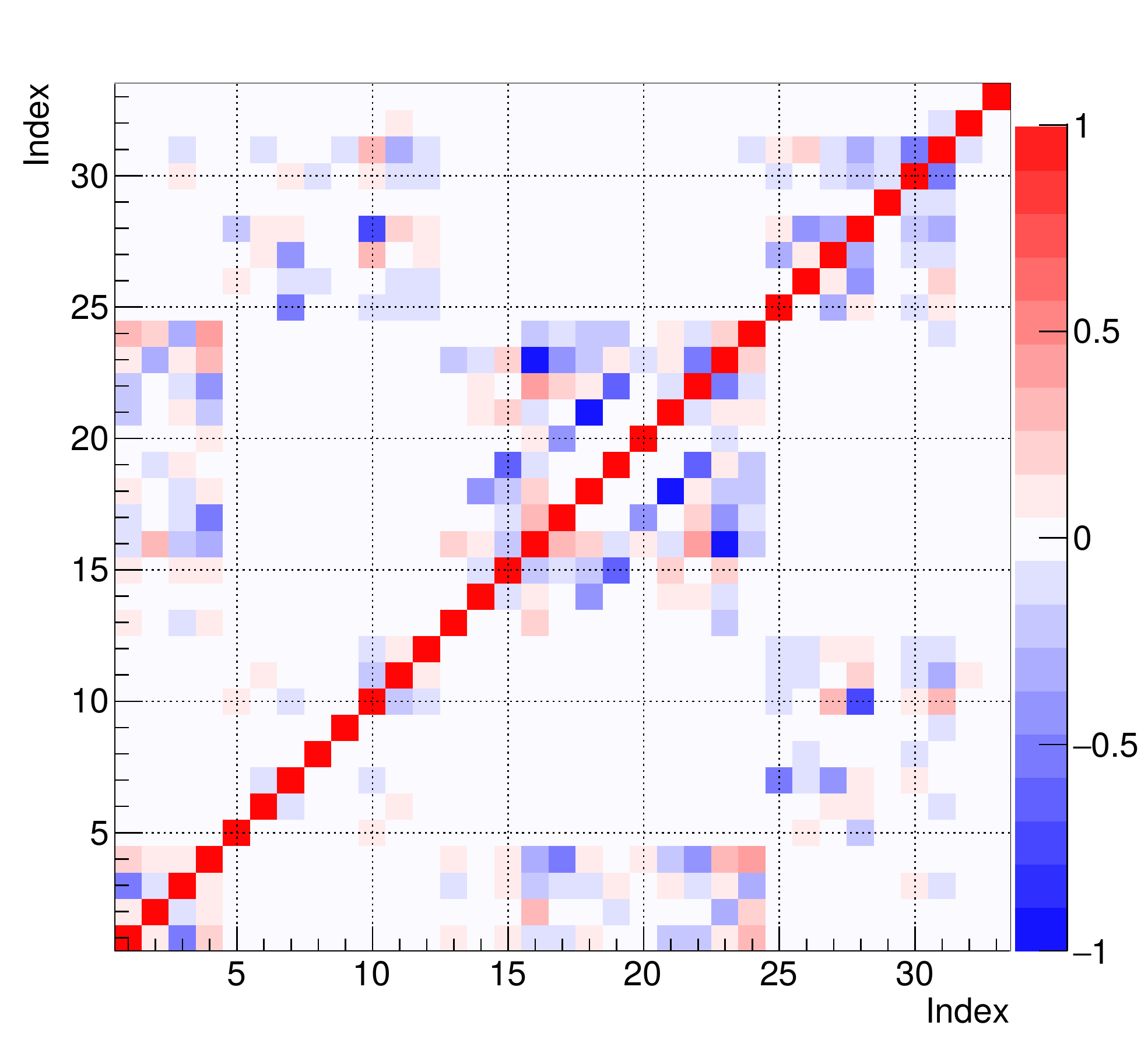}
\caption{Correlation matrix among the reconstructed activities. The source indexes are from Table~\ref{tab:Results}.}
\label{fig:CorrMatrix}
\end{center}
\end{figure}

We perform a simultaneous Bayesian fit of \monealfa, \monebeta, \mtwo and \mtwosum spectra with a linear combination of 33 background sources, to evaluate their activities. We use the JAGS (Just Another Gibbs Sampler) software~\cite{JAGS} to define the Bayesian statistical model and to sample the \textit{joint posterior} Probability Density Function (PDF) of the fit parameters (i.e. the normalization coefficients of the Monte Carlo spectra). More details about the JAGS-based analysis tool for background model fit can be found in Ref.~\cite{Alduino:2016vtd_Q02nbb}. We use non-negative uniform priors for all fit parameters, with a few exceptions discussed in Section~\ref{sec:BkgModel}.

We choose a variable step size binning to minimize the effect of not ideal detector response which manifests itself in line shapes of complicated modeling, especially in the $\alpha$ region. Practically, we define a binning that does not split the peaks in more than one bin. We set the minimum bin step to 15~keV in \monebeta spectrum and to 25~keV in \mtwo and \mtwosum spectra. We enlarge the binning in the regions with low density of events, in order to minimize the effects of statistical fluctuations.

The fit range extends from 300~keV to 5~MeV, and from 2~MeV to 11~MeV for \monebeta and \monealfa spectrum, respectively. The multiplicity 2 events used to fill \mtwo and \mtwosum spectra are selected by requiring that both events in the multiplet have energy above a threshold set to 150~keV. 
Excluding the low energy events allows to bypass problems related to noise pulses that sometimes can be triggered and that are difficult to be discriminated from low energy physics events.

We label as \textit{reference} the fit performed with the binning and energy range described in this section, using the sources of the reference background model. The effect of different choices is investigated in the systematic studies.
The results of the \textit{reference} fit to the experimental data collected with a \exposure \znsenr exposure are shown in Figures \ref{fig:JAGSM1} and \ref{fig:JAGSM2}. 
In these plots, we show the comparison between the experimental and the fit-reconstructed spectra.
The pull distribution, obtained from the fit residuals of all bins, is shown in Figure~\ref{fig:Pulls} and is compatible with a Gaussian with $\mu=0$ and $\sigma=1$.

We analyze the marginal posterior distributions of the fit parameters to evaluate the activities of background sources. Most of the marginal PDFs have a Gaussian shape and we calculate their mean and standard deviation to get the activity and its uncertainty. Differently, when the activity of a source is compatible with zero, we quote a 90\% upper limit by integrating the posterior PDF.

From the sampling of the \textit{joint posterior} PDF, we also extract the correlation matrix between the fit parameters, represented in Figure~\ref{fig:CorrMatrix}. As expected, the \textit{internal/near} sources used to fit the \monealfa spectrum are not correlated to the others, while the sources representing the same contaminant in different positions of the cryostat are highly (anti-)correlated.

The activities of the sources used in the \textit{reference} fit are listed in Table~\ref{tab:Results}.
These numbers must be read and interpreted keeping in mind the approximations and the choices made in constructing the background model. 
Particularly, since the fit is performed on the full statistics collected by all \cuz detectors, the activities evaluated for \crystals and \reflectors are average values of real contaminations, that could be not uniformly distributed.
Moreover, the representation of the external sources is extremely simplified and does not aspire to establish with accuracy the activities of sources in cryostat and shields. 
Despite the above caveats, this method allows to constrain the background sources on their specific signatures in the experimental data, and to extrapolate their contribution to the \bbz ROI on a relative scale, independently of the absolute activity evaluated for each source.

\subsection{Analysis of \bbz ROI and systematics}

\begin{figure*}[!htb]
\begin{center}
\includegraphics[width=.99\textwidth]{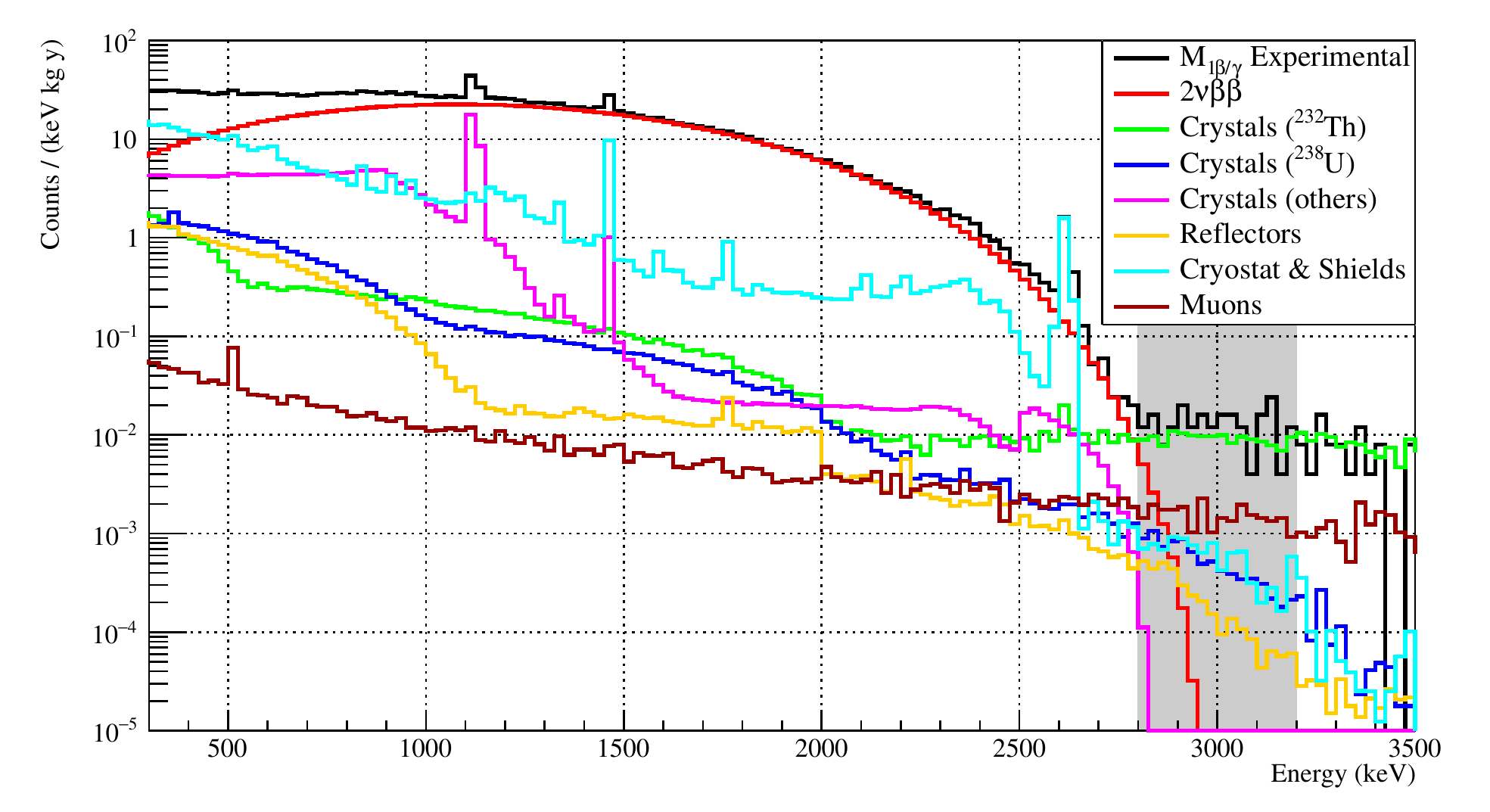}
\caption{Background sources contributing to the \monebeta reconstruction, grouped by source and component. The shaded area corresponds to the 400~keV energy range from 2.8~MeV to 3.2~MeV (ROI$_{bkg}$) chosen to analyze the background in the region of interest around the \seod \bbz Q-value. In this plot, the time veto for the rejection of $^{208}$Tl events is not applied, thus the ROI$_{bkg}$ is dominated by the $\beta/\gamma$-events from \th chain contaminations located in \crystals.}
\label{fig:PlotComponents}
\end{center}
\end{figure*}

To analyze the background in the region of interest around the \seod \bbz Q-value, we define a 400~keV interval from 2.8~MeV to 3.2~MeV (hereinafter referred to as ROI$_{bkg}$). The energy range chosen for ROI$_{bkg}$ is much wider with respect to the ($23\pm0.6$)~keV FWHM energy resolution at Q-value~\cite{Azzolini:2018dyb_C0PRL}, in order to include enough experimental counts to be used as a benchmark for background model predictions.
In Figure~\ref{fig:PlotComponents}, we show the reconstructed \monebeta spectra of different groups of sources, and their contribution to the ROI$_{bkg}$ counting rate.

The counts predicted by the background model in the ROI$_{bkg}$ are 50.5~$\pm$~1.3. This is perfectly compatible with the 52 counts experimentally observed.
Most part of ROI$_{bkg}$ events are due to $^{208}$Tl decays inside the \crystals. This radioisotope belongs to the lower part of \th chain and decays via $\beta/\gamma$ with a Q-value of about 5~MeV.
Based on the background model results, the $^{228}$Ra--$^{208}$Pb source in the bulk of \crystals produces $(8.4\pm0.3)\times10^{-3}$\rate. The same contaminant on the surfaces of \crystals results in a rate of $(7.8\pm1.2)\times10^{-4}$\rate. These rates correspond to a total amount of $\sim$37 counts in the ROI$_{bkg}$.
Nevertheless, since the $^{208}$Tl half-life is relatively short (3.05~min), the $^{208}$Tl events can be rejected by exploiting the time coincidence with its parent, $^{212}$Bi~\cite{Azzolini:2018yye_C0Analysis}. For each event in the ROI$_{bkg}$, we check if it is preceded by a $^{212}$Bi-like $\alpha$ event in the same crystal.
By applying a 7$\times$\vita-dim window time veto to the simulations of $^{228}$Ra--$^{208}$Pb sources, the predicted counting rates in the ROI$_{bkg}$ become $(3.4\pm0.6)\times10^{-4}$\rate and $(3.4\pm0.5)\times10^{-4}$ \rate for bulk and surface contaminations of \crystals, respectively. Hence the model predicts the rejection of 34~$\pm$~1 counts. In the experimental data we observe the rejection of 38 events. We ascribe the higher event rejection in the experimental data to random coincidences in the time veto window.   

Nonetheless, after applying the time veto, the background model predicts 16.5~$\pm$~0.8 counts in ROI, which is still well compatible with the 14 measured ones. 

In Table~\ref{tab:ROI}, we report the counting rates reconstructed in the ROI$_{bkg}$ through the background model for the different sources after applying the time veto to reject $^{208}$Tl events. 
We sum the contributions from surface contaminations at different depths, and from the different components of cryostat and shields, thus reducing the effect of anti-correlation affecting these sources.
As shown in Figure~\ref{fig:PlotComponents}, the spectra of background sources in the ROI$_{bkg}$ exhibit either a flat or a decreasing trend, with no significant peaking structures. 
Since the ROI$_{bkg}$ is symmetrical around \seod \bb Q-value, the counting rate in the ROI$_{bkg}$ is a good approximation of the expected rate in the narrower region where the \bbz signature is searched. This is true for all background components except for the \bbd one, because its spectrum has the endpoint at the Q-value of \seod \bb decay. The contribution from \bbd source reported in Table~\ref{tab:ROI} is produced exclusively by events with energy $<2950$~keV, while the expected counting rate from \bbd in a 100~keV range centered at \seod Q-value is $<3\times10^{-6}$\rate.

In order to study the systematic uncertainties of the background reconstruction in the ROI$_{bkg}$, we perform some fits in which the sources are modeled in a different way with respect to the \textit{reference} fit. 
Particularly, we performed the following tests:
\begin{enumerate}
\item a fit with a \textit{reduced list} of sources in which we exclude the contaminations evaluated as upper limits in the \textit{reference} fit;
\item a fit with \crystals surface contaminations simulated by setting the depth parameter at 0.1~$\mu$m instead of 0.01~$\mu$m ;
\item a fit in which the $^{226}$Ra--$^{210}$Pb contamination in \reflectors is removed from the list of sources;
\item a fit in which the \reflectors sources simulated with 10~$\mu$m depth parameter are replaced by uniformly distributed contaminations;
\item a fit in which we add \th and \u contaminations on \holder surfaces ($\lambda=10\mu$m), constrained by priors from \qz background model~\cite{Alduino:2016vtd_Q02nbb};
\item a fit in which we investigate the effect of \th and \u surface contaminations on the 50~mK shield surrounding the \cuz tower;
\item three fits in which the source list does not include \th and \u contaminations in \cryoint, \pbint, and \pbext, respectively;
\end{enumerate}

\renewcommand{\arraystretch}{1.3}
\begin{table*}[!t]
\begin{center}
\caption{Counting rates reconstructed in the ROI$_{bkg}$ (from 2.8~MeV to 3.2~MeV) for the different sources, after applying the time veto for the rejection of $^{208}$Tl events. For each value of counting rate we quote first the statistical uncertainty and then the systematic one. In the left column we report the total counting rates from the different components of the experimental setup, while in the right column we provide their breakdown by source. \th and \u refers the chain parts producing background in the ROI$_{bkg}$, i.e $^{228}$Ra--$^{208}$Pb and $^{226}$Ra--$^{210}$Pb, respectively. The counting rate quoted for \reflectors includes also the contribution from surface contaminations of the \holder. The \th source limit (90\% C.L.) corresponds to the maximum one obtained from the analysis of systematic uncertainties. The contribution from \bbd is produced exclusively by events with energy $<2950$~keV. }
\begin{tabular}{ll|ll}
\hline
Component & ROI$_{bkg}$ rate & Source & ROI$_{bkg}$ rate \\
          & ($10^{-4}$\rate) &        & ($10^{-4}$\rate) \\
\hline
\multirow{3}{*}{\crystals} &  \multirow{3}{*}{$11.7\pm0.6$ $_{-0.8}^{+1.6}$  }& \th -- bulk &	$3.4\pm0.6\pm0.1$		\\
                           &                                 & \th --surf  &    $3.4\pm0.5$ $_{-0.7}^{+1.0}$ $ $	\\
                           &                                 & \u --surf   &	$4.9\pm0.3$ $_{-0.3}^{+1.3}$ $ $	\\
\hline
\multirow{2}{*}{\textit{Reflectors} \& \textit{Holder}} &  \multirow{2}{*}{$2.1\pm0.3$ $_{-1.0}^{+2.2}$   } & \th 	&	$<3.3$ \\
                           &                                 &   \u 	&	$1.8\pm0.3$ $_{-0.9}^{+1.4}$ $ $	  \\ 
\hline
\multirow{2}{*}{\textit{Cryostat} \& \textit{Shields}} &  \multirow{2}{*}{$5.9\pm1.3$ $_{-2.9}^{+7.2}$   } & \th 	&	$3.5\pm1.3$ $_{-3.3}^{+7.4}$ $ $	 \\
                           &         &   \u 	&	$2.4\pm0.4$ $_{-0.7}^{+4.1}$ $ $	 \\
\hline
\textbf{Subtotal} & \multicolumn{3}{l}{$19.8\pm1.4$ $_{-2.7}^{+6.6}$ } \\
Muons		    &	\multicolumn{3}{l}{$15.3\pm1.3\pm2.5$}		\\
\bbd           	&	\multicolumn{3}{l}{$6.0\pm0.3$ ($<3\times10^{-6}$ \rate in [2.95--3.05]~MeV range) } \\
\hline
\textbf{Total}			&	\multicolumn{3}{l}{$41\pm2$ $_{-4}^{+9}$ 	}		 \\
\hline
\textbf{Experimental}			&	\multicolumn{3}{l}{$35$ $_{-9}^{+10}$	}		 \\
\hline
\end{tabular}

\label{tab:ROI}
\end{center}
\end{table*}

In all of these tests, we obtain pull distributions compatible with a standard Gaussian. Therefore, we analyze the differences in the ROI$_{bkg}$ counting rates to get an estimate of systematic uncertainties, reported in Table~\ref{tab:ROI}.
We do not quote a systematic uncertainty for \bbd contribution to the ROI$_{bkg}$, because the results from all tests are within a range much smaller than the statistical uncertainty. 
\crystals surface contaminations are constrained by the time analysis of consecutive $\alpha$ decays. Their counting rate in the ROI$_{bkg}$ has a maximum variation of $\sim$30\% when fitting with the \textit{reduced list} (that does not include 10~$\mu$m surface contaminations of \crystals) and when performing the tests number 2 and 3.

The systematic uncertainties affecting the ROI$_{bkg}$ counting rate due to \reflectors and \holder contaminations are investigated through tests number 3, 4, and 5. 
The \textit{bulk}/\textit{deep surface} contaminations in \reflectors produce a continuum of degraded $\alpha$ that allows to obtain a good fit to the \monealfa spectrum in the [2--4]~MeV range. 
Since \th in \reflectors is constrained by a prior which makes negligible its contribution, $^{226}$Ra--$^{210}$Pb and $^{210}$Pb--$^{206}$Pb are the only reflector sources which are left free to fit this continuum. 
In fit number 3, we investigate the scenario in which \reflectors are contaminated only by $^{210}$Pb--$^{206}$Pb. The result is that the experimental counts in the [7--7.5]~MeV range of \monealfa spectrum are not reconstructed by the model and we get a 5$\sigma$ disagreement in that bin. 
We conclude that a contribution from the $^{226}$Ra--$^{210}$Pb source in \reflectors is needed to preserve the fit quality and we estimate that its activity (and thus its counting rate in the ROI) must be at least half of that evaluated in the \textit{reference} fit.
On the other hand, the result of fit number 4, in which we choose an equally plausible model for the distribution of contaminants in \reflectors is that the ROI$_{bkg}$ counting rate from this source increases by $\sim$50\%.
The fit number 5 is aimed at investigating the effect of contaminations on \holder surfaces. This background source does not have a specific signature in the experimental data, because most of the $\alpha$ particles generated at \holder surfaces are absorbed by \reflectors. However, some $\beta$ particles from \u and \th chains can cross the reflector foils and produce events in the ROI$_{bkg}$. Since the \holder is made of the same NOSV copper used in \qz, we exploit the values of \th and \u surface contaminations measured with \qz detector to constrain these sources. The result is that the reconstructed rate in the ROI$_{bkg}$ increases by $\sim1.5\times10^{-4}$~\rate. Particularly, the upper limit on the background due to \th in \reflectors and \holder becomes less stringent: $<3.3\times10^{-4}$\rate. 
Similarly, in test number 6 we evaluate the systematic uncertainty affecting the ROI$_{bkg}$ reconstruction in the case that \th and \u surface contaminations on the 50~mK shield surrounding \cuz tower are not negligible with respect to the bulk contaminations of \cryoint. Also in this case, as expected, the fit predicts a higher rate in the ROI$_{bkg}$, with an increase of $\sim4\times10^{-4}$~\rate with respect to the \textit{reference} one.
The fits of test number 7 are used to investigate how the uncertainty on location and description of sources in cryostat and shields is propagated to the estimate of their contribution to the ROI$_{bkg}$.

Finally, we performed some tests in which we varied the minimum bin size (from 5 to 50 keV) and the energy calibration (according to the residuals of a $^{56}$Co calibration measurement reported in Ref.~\cite{Azzolini:2018yye_C0Analysis}). The changes in source activities and ROI$_{bkg}$ reconstruction are much smaller than the uncertainties quoted in Tables \ref{tab:Results} and \ref{tab:ROI}, thus the corresponding systematic errors are negligible.

\section{Conclusion and perspectives}
In this paper we fit the \cuz data using 33 radioactive sources, modeled via Monte Carlo simulations. 
We identify the contribution of the various background sources to the ROI$_{bkg}$ counting rate and we perform an analysis of the corresponding systematic uncertainties. Excluding the \bbd decay contribution (which is negligible at the Q-value of \seod \bb decay), we obtain that $\sim$44\% of background rate in the ROI$_{bkg}$ is produced by cosmic muon showers, while the remaining fraction is due to radioactive decays in \crystals ($\sim$33\%), in \reflectors \& \holder ($\sim$6\%), and in \textit{Cryostat} \& \textit{Shields} ($\sim$17\%).

Based on these results, an upgrade of the \cuz detector has been scheduled in order to reduce the background level in the ROI and to further improve the comprehension of background sources.
In \cuz Phase-II we plan to install a muon veto, which will be implemented through a system of plastic scintillators in the external experimental setup. 
Moreover we will investigate the effect of removing reflecting foils. This will also allow us to get more information about surface contaminations of \crystals, through the analysis of \mtwo and \mtwosum spectra, as performed in CUORE-0~\cite{Alduino:2016vjd_Q0det}.
Finally, we will add an ultra-pure copper vessel at 10~mK, acting as thermal and radioactive shield, that is expected to further reduce the counting rate due to contaminations of cryostat and shields.

The \cuz results on $\alpha$-background rejection, further strengthened by the analysis presented here, are a founding pillar of the next-generation CUPID experiment~\cite{Wang:2015raa,Wang:2015taa}, based on scintillating calorimeters.

\section{Acknowledgments}
This work was partially supported by the European Research Council (FP7/2007-2013) under contract LUCIFER no. 247115.
We thank M. Iannone for his help in all the stages of the detector assembly, A. Pelosi for constructing the assembly line, M. Guetti for the assistance in the cryogenic operations, R. Gaigher for the mechanics of the calibration system, M. Lindozzi for the cryostat monitoring system, M. Perego for his invaluable help in many tasks, the mechanical workshop of LNGS (E. Tatananni, A. Rotilio, A. Corsi, and B. Romualdi) for the continuous help in the overall set-up design.
We acknowledge the Dark Side Collaboration for the use of the low-radon clean room.
This work makes use of the DIANA data analysis and APOLLO data acquisition software which has been developed by the CUORICINO, CUORE, LUCIFER and \cuz collaborations.
This work makes use of the Arby software for Geant4 based Monte Carlo simulations, that has been developed in the framework of the Milano -- Bicocca R\&D activities and that is maintained by O. Cremonesi and S. Pozzi.
\bibliography{main}  % name your BibTeX data base

\begin{thebibliography}{10}
\providecommand{\url}[1]{{#1}}
\providecommand{\urlprefix}{URL }
\expandafter\ifx\csname urlstyle\endcsname\relax
  \providecommand{\doi}[1]{DOI \discretionary{}{}{}#1}\else
  \providecommand{\doi}{DOI \discretionary{}{}{}\begingroup
  \urlstyle{rm}\Url}\fi

\bibitem{Furry:1939qr}
W.H. Furry, Phys. Rev. \textbf{56}, 1184 (1939).
\newblock \doi{10.1103/PhysRev.56.1184}

\bibitem{DellOro:2016tmg}
S.~Dell'Oro, S.~Marcocci, M.~Viel, F.~Vissani, Adv. High Energy Phys.
  \textbf{2016}, 2162659 (2016).
\newblock \doi{10.1155/2016/2162659}

\bibitem{Alduino:2017pni_Sensitivity}
C.~Alduino, et~al., Eur. Phys. J. \textbf{C77}(8), 532 (2017).
\newblock \doi{10.1140/epjc/s10052-017-5098-9}

\bibitem{Lincoln:2012fq}
D.L. Lincoln, J.D. Holt, G.~Bollen, M.~Brodeur, S.~Bustabad, J.~Engel, S.J.
  Novario, M.~Redshaw, R.~Ringle, S.~Schwarz, Phys. Rev. Lett. \textbf{110},
  012501 (2013).
\newblock \doi{10.1103/PhysRevLett.110.012501}

\bibitem{Azzolini:2018dyb_C0PRL}
O.~Azzolini, et~al., Phys. Rev. Lett. \textbf{120}(23), 232502 (2018).
\newblock \doi{10.1103/PhysRevLett.120.232502}

\bibitem{Andreotti:2010vj}
E.~Andreotti, et~al., Astropart. Phys. \textbf{34}, 822 (2011).
\newblock \doi{10.1016/j.astropartphys.2011.02.002}

\bibitem{Alfonso:2015wka_Q0PRL}
K.~Alfonso, et~al., Phys. Rev. Lett. \textbf{115}(10), 102502 (2015).
\newblock \doi{10.1103/PhysRevLett.115.102502}

\bibitem{Beeman:2015xjv}
J.W. Beeman, et~al., Eur. Phys. J. \textbf{C75}(12), 591 (2015).
\newblock \doi{10.1140/epjc/s10052-015-3822-x}

\bibitem{Azzolini:2018tum_C0det}
O.~Azzolini, et~al., Eur. Phys. J. \textbf{C78}(5), 428 (2018).
\newblock \doi{10.1140/epjc/s10052-018-5896-8}

\bibitem{Alduino:2016vtd_Q02nbb}
C.~Alduino, et~al., Eur. Phys. J. \textbf{C77}(1), 13 (2017).
\newblock \doi{10.1140/epjc/s10052-016-4498-6}

\bibitem{Alduino:2016vjd_Q0det}
C.~Alduino, et~al., JINST \textbf{11}(07), P07009 (2016).
\newblock \doi{10.1088/1748-0221/11/07/P07009}

\bibitem{BiPoDetector}
A.S. Barabash, et~al., JINST \textbf{12}(06), P06002 (2017).
\newblock \doi{10.1088/1748-0221/12/06/P06002}

\bibitem{Wang:1989vk}
N.~Wang, F.C. Wellstood, B.~Sadoulet, E.E. Haller, J.~Beeman, Phys. Rev.
  \textbf{B41}, 3761 (1990).
\newblock \doi{10.1103/PhysRevB.41.3761}

\bibitem{Andreotti:2012zz}
E.~Andreotti, et~al., Nucl. Instrum. Meth. A \textbf{664}, 161 (2012).
\newblock \doi{10.1016/j.nima.2011.10.065}

\bibitem{Carniti:2017zkr}
K.~Alfonso, L.~Cassina, A.~Giachero, C.~Gotti, G.~Pessina, P.~Carniti, JINST
  \textbf{13}(02), P02029 (2018).
\newblock \doi{10.1088/1748-0221/13/02/P02029}

\bibitem{Arnaboldi:2015wvc}
C.~Arnaboldi, et~al., Rev. Sci. Instrum. \textbf{86}(12), 124703 (2015).
\newblock \doi{10.1063/1.4936269}

\bibitem{Arnaboldi:2017aek}
C.~Arnaboldi, P.~Carniti, L.~Cassina, C.~Gotti, X.~Liu, M.~Maino, G.~Pessina,
  C.~Rosenfeld, B.X. Zhu, JINST \textbf{13}(02), P02026 (2018).
\newblock \doi{10.1088/1748-0221/13/02/P02026}

\bibitem{DiDomizio:2018ldc}
S.D. Domizio, A.~Branca, A.~Caminata, L.~Canonica, S.~Copello, A.~Giachero,
  E.~Guardincerri, L.~Marini, M.~Pallavicini, M.~Vignati, Journal of
  Instrumentation \textbf{13}(12), P12003 (2018).
\newblock \doi{10.1088/1748-0221/13/12/p12003}

\bibitem{Alduino:2016zrl_Q0Analysis}
C.~Alduino, et~al., Phys. Rev. \textbf{C93}(4), 045503 (2016).
\newblock \doi{10.1103/PhysRevC.93.045503}

\bibitem{Azzolini:2018yye_C0Analysis}
O.~Azzolini, et~al., Eur. Phys. J. \textbf{C78}(9), 734 (2018).
\newblock \doi{10.1140/epjc/s10052-018-6202-5}

\bibitem{Azzolini:2018oph_C0ExcitedStates}
O.~Azzolini, et~al., Eur. Phys. J. \textbf{C78}(11), 888 (2018).
\newblock \doi{10.1140/epjc/s10052-018-6340-9}

\bibitem{Gatti:1986cw}
E.~Gatti, P.F. Manfredi, Riv. Nuovo Cimento \textbf{9}, 1 (1986)

\bibitem{Radeka:1966}
V.~Radeka, N.~Karlovac, Nucl. Instrum. Meth. \textbf{52}, 86 (1967)

\bibitem{Artusa:2016maw}
D.R. Artusa, et~al., Eur. Phys. J. C \textbf{76}(7), 364 (2016).
\newblock \doi{10.1140/epjc/s10052-016-4223-5}

\bibitem{Alduino:2017qet_BkgBudget}
C.~Alduino, et~al., Eur. Phys. J. C \textbf{77}(8), 543 (2017).
\newblock \doi{10.1140/epjc/s10052-017-5080-6}

\bibitem{Arnold:2018tmo}
R.~Arnold, et~al., Eur. Phys. J. \textbf{C78}(10), 821 (2018).
\newblock \doi{10.1140/epjc/s10052-018-6295-x}

\bibitem{Sm147}
N.~Nica, Nuclear Data Sheets \textbf{110}(4), 749  (2009).
\newblock \doi{https://doi.org/10.1016/j.nds.2009.02.003}

\bibitem{Geant4}
S.~Agostinelli, {\it et al.}, Nucl. Instr. Meth. A \textbf{506}, 250 (2003)

\bibitem{Kotila:2012zza}
J.~Kotila, F.~Iachello, Phys. Rev. \textbf{C85}, 034316 (2012).
\newblock \doi{10.1103/PhysRevC.85.034316}

\bibitem{Andreotti:2009dk}
E.~Andreotti, et~al., Astropart. Phys. \textbf{34}, 18 (2010).
\newblock \doi{10.1016/j.astropartphys.2010.04.004}

\bibitem{Alessandria:2012zp}
F.~Alessandria, et~al., Astropart. Phys. \textbf{45}, 13 (2013).
\newblock \doi{10.1016/j.astropartphys.2013.02.005}

\bibitem{Ambrosio:1995cx}
M.~Ambrosio, et~al., Phys. Rev. \textbf{D52}, 3793 (1995).
\newblock \doi{10.1103/PhysRevD.52.3793}

\bibitem{JAGS}
M.~Plummer, JAGS: A Program for Analysis of Bayesian Graphical Models using
  Gibbs Sampling. 3rd International Workshop on Distributed Statistical
  Computing (DSC 2003); Vienna, Austria. \textbf{124} (2003)

\bibitem{Wang:2015raa}
{CUPID Interest group}, arXiv:1504.03599  (2015).
\newblock \urlprefix\url{https://arxiv.org/abs/1504.03599}

\bibitem{Wang:2015taa}
{CUPID Interest group}, arXiv:1504.03612  (2015).
\newblock \urlprefix\url{https://arxiv.org/abs/1504.03612}

\end{thebibliography}
\bibliographystyle{spphys} 
\end{document}